# ゲームジャンルのゲーム仕様の系譜を可視化する―育成ゲームジャンルについての事例研究を通して―

井上明人, 毛利仁美

立命館大学映像学部, 立命館大学ゲーム研究センター

## 1.はじめに

　ビデオゲームにおける「ジャンル」の系譜はビデオゲーム研究における重要な課題の一つとなっている。

　特定の作品を論じるのではなく、何かしらの一群のカテゴリーを問題にしようと思ったとき、ビデオゲームの「ジャンル」はそのカテゴリーを示すものとして頻繁に用いられ、かつ意味のある概念であるとされる(Goddard & Muscat 2017)。ごく基礎的な概念であるがゆえに、この概念の運用はビデオゲームに関わる経済的研究、文化的研究そして実制作などの多様な側面においてしばしばクリティカルな意味をもつ。しかし、このように重要な概念であるにも関わらず、ビデオゲームのジャンルをどのように同定するのか、そのジャンルの特徴は何か、作品間の関係がジャンル内でどのようになっているのかといった基本的な点を示す効果的な手法はまだ発展途上である。

　ただし、ビデオゲームの「ジャンル」概念自体の研究は存在している。ゲームの「ジャンル」概念は、Wolf(2002)によるジャンル区分の提案や、Lee et al.(2014)によるファセットの提案、Lessard(2014)が示すように、ゲームのメカニクス（ルールやアルゴリズム等）といった層に加えて、ゲームが表現しようとする主題（中世ヨーロッパにおける戦争、現代の日本の女の子のファッション）といった多層的な要素を含んでいる。後者のようなジャンルの性質は、小説や映画等の表象メディアでも見られるジャンルの概念であるが、特にビデオゲームにおける特徴的な要素としては前者のメカニクスの点であるとされる(Arsenault 2009, p.155)。

本研究の目標の第一に、特にビデオゲームに固有の特徴的要素であるゲームメカニクスの系譜を明示的に示すために手法を提案・開発する。また第二に、具体的な手法の事例研究として「育成ゲーム」ジャンルの系譜を提案した手法を通じて可視化するものである。なお、本稿では、ゲームメカニクスのことを、日本のゲーム業界の業界慣行で用いられる用語に寄せて、以後「ゲーム仕様」として記述する。

## 2.先行研究

### 2.1 ゲームジャンルとは何か：形式、言説

　ゲーム仕様を基盤とした系譜図を描くにあたって、さまざまな研究があるが、ゲームジャンルに関する研究は、未だジャンル概念の扱いやジャンル内の作品の関係性などを示す方法について、確固たる手法が確立されているというわけではない。

　ゲームジャンルの系譜の可視化を考えた場合、大きな論点としては、第一に、どのように特定のジャンルの範囲を同定するのかということと、第二に、どのように系譜の関係を示すか、という２点について大きく集約される。これらについて、どのような議論がなされてきたかについて、まず概観した上で、本調査研究の方法論を検討していきたい。

### 2.2 特定のゲーム仕様のジャンルの範囲をどう同定するか

　第一にジャンルをどう同定するかという点であるが、これについてはそもそもゲームジャンルとは何かという概念が最初に問題となる。

　Clearwater(2011)が指摘するように、ある特定のジャンル概念が示す範囲とは、技術的な要因のみによって決定されるものではない。様々な社会的な言説や産業的な構造等に影響されながら作られるものである。ジャンルの成立・変化のプロセスは、ゲームのルールなどのアイデアや、コンピュータのソフトやハード、ネットワーク等の技術のみならず、雑誌やウェブの言説等といった多様な影響のなかで成立している。そして、その意味範囲は継続的に変化するプロセスの中におかれ、概念が完全に安定するような性質のものではない（Wolf 2002; Apperly 2006; Arsenault 2009; Clearwater 2011）。こうした曖昧な特性を持つものであるがゆえに、ビデオゲームのジャンルがどのように同定できるのか、あるいはそのジャンルの中のサブジャンルの区別がどうなっているのか決定することは困難を伴う（Juul 2016）。

すなわち、技術や設計等の観点から、論理的に相互排他的なジャンルの同定を行うことはそもそも難しい性質のものであると言える。

この論点に対して Goddard & Muscat(2017)はゲーム・デザインの見地からゲームのジャンルを検討、分析するための方法として、そのジャンルに類する個別の作品を挙げたリストを収集するという方法を提案している。つまり、ジャンルの概念が社会的に作られるという側面が強いものであるならば、雑誌や、ウェブの言説や、特定ジャンルのリストとして挙げられたものを収集することで、あるゲームのジャンルの大まかな範囲を把握することができるだろうというものである。

## 2.3 系譜図を可視化する方法論をめぐって

系譜図の提案という点で特に実践的な例を提示しているものとしては、Zagal et al.(2005)によるオントロジーの提案や、Juul(2016)によるゲームの系譜図の提案であろう。Juul(2016)は、パズルゲームのうちのサブジャンルである「マッチングタイルゲーム」の系譜図を、主要なゲームがどのようにゲーム仕様を変化させてきたか、という点に注目しながらそのネットワーク構造を描画している（図 2-1）。

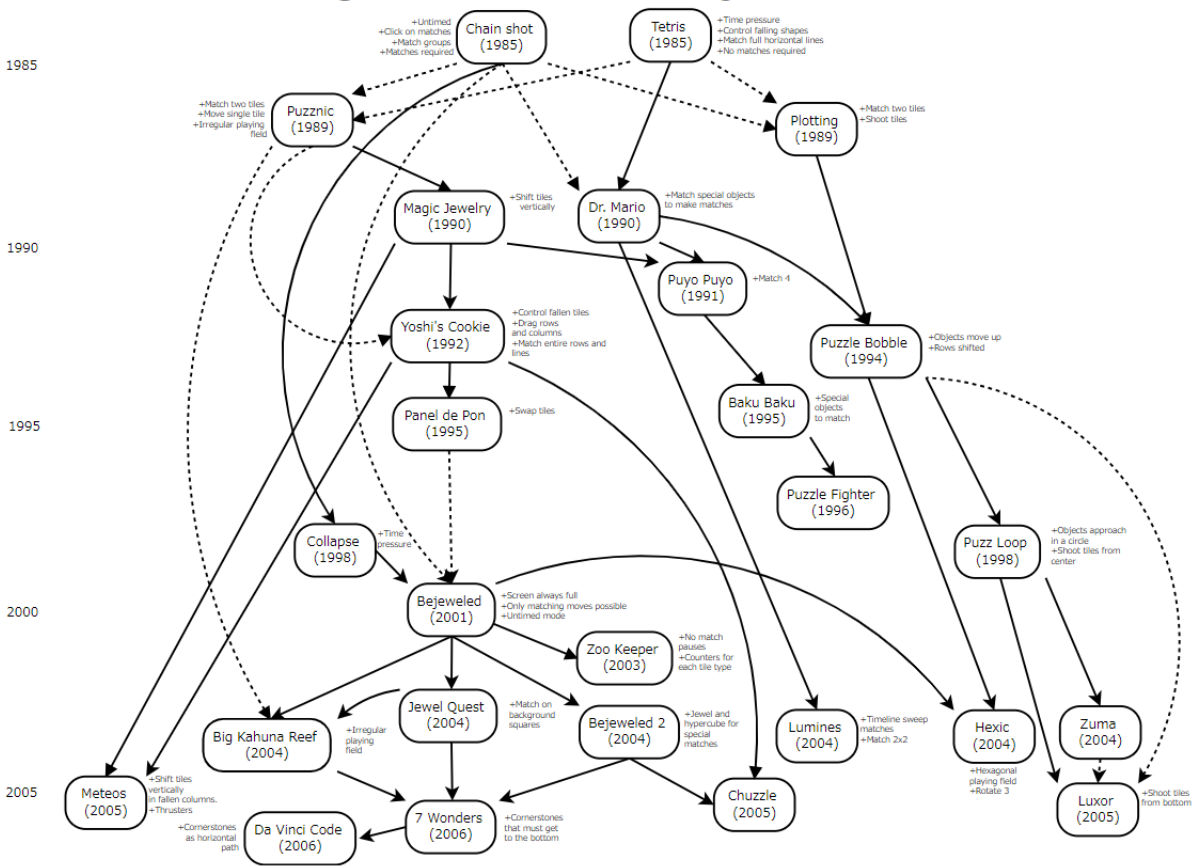

図 2-1：Juul (2016)によるマッチングタイルゲームの系譜図

　特定のゲーム仕様の変化を土台に、ゲームジャンルの系譜図を描くかという方法は、個別具体的に、作品同士の関係性を第三者が検討可能であるという点で、説得的な手法であると言えるだろう。

　この点については、ゲームジャンル概念についての包括的な議論を行っている Goddard & Muscat(2017)も、ジャンルの範囲の同定ののち、その特定のジャンルを理解するための体系的かつ分析的な調査のための、詳細な要素を挙げることを提案している。そして、そこで挙げられる要素には、「明確に表現され、反論可能なものとして表現されている必要」があるとしている。

## 3.方法論

　以上の先行研究を踏まえ、本研究では可能な限り再現性の高い手法を用いて、ゲームジャンルをめぐる社会的な言説、ゲーム仕様の双方の観点から特定ジャンルについての分析を行いたい。

　具体的には、様々なゲームジャンルのうちから「育成ゲーム」を選定して、そのゲームジャンル内の系譜を明らかにしていく。育成ゲームが本調査手法を用いる対象として適切である理由は二点ある。

　第一に、特に重要とされる主要な作品を実際にプレイして内容を確認して示すことが、相対的に可能であろうと思われるためである。ロールプレイングゲーム（以降、RPG）やシミュレーションゲーム、アドベンチャーゲーム、といったジャンルでは重要とされる作品を選定し、内容確認を行うだけでも膨大な時間を必要とされるものと想定されるが、育成ゲームについては重要作品の量がそれらのジャンルに比べると相対的に少ないと思われ、調査の作業が実施可能であると思われる。

　第二に、ジャンルの概念自体がある程度安定しつつあると思われることである。4.1 で後述するとおり「育成ゲーム」をめぐる言説の量自体が 2000 年代以後に減少してきており、ジャンルの概念としてある程度安定してきていると思われるためである。

　調査の方法論として、先行研究において議論されている論点と、提案されている手法を勘案し、次のような手法を採る。

(1) 雑誌・メディア言説から「育成ゲーム」の範囲を検討する（Goddard & Muscat 2017）

　　　「育成ゲーム」言説のなかで数多く言及されていたゲームを対象に、内容を確認すべきゲームを選定する。(4.1.1〜4.1.6)

(2) その上で、対象となるゲームを選定する。
　　　なお、新しい作品についてはある作品が「育成ゲーム」の系譜に加わるかどうかの評価言説が安定していない可能性があることを考え 2020 年以前の言説に限定する。
　　（4.1.7）

(3) 選定したゲームの内容を確認し、一つのゲームにつきおよそ 50 個前後のゲーム仕様を列挙する（Goddard & Muscat 2017）

  系統誤差を防ぐため、複数人がゲームの内容を確認する[1]。ゲームの内容の確認にあたっては、ビデオゲームの専門の開発者およびビデオゲームの研究を専門とする大学院生、ビデオゲーム分野の開発についての教育課程を持つ大学の学部生が内容の確認を実施する。この作業はすべてのゲームについて 2 名以上で実施する。(4.2 以後の元データの作成)

(4) 確認した内容について、2 名でのゲーム仕様の粒度のすり合わせを行う(川喜田 1970)

  カテゴリーの粒度について合意を得るための手法として、KJ 法(川喜田 1970)を用いる。

  本調査は内容分析に近い側面をもっており内容分析の調査手法を参考とした。内容分析にあたって、各調査者の理解度の一致率（信頼性）を確認するための方法として、複数の調査者による内容の一致度を確認することが推奨されている(有馬 2007)。(4.2 以後の元データの作成)

(5)すり合わせをおこなったのち、各ゲームの要素が相互にどのように共通しているかを洗い出す。この処理は二段階で実施する。(4.2 以後の元データの作成)

(5-1)ゲーム仕様の記述文の完全一致を元とした処理：

  ある作品のゲーム仕様の記述文と同一の文言のゲーム仕様の記述文が別の作品に見られた場合に、双方のゲームに共通するゲーム仕様が一点あるとカウントする[2]。たとえば「ゲームの最初にプレイヤーキャラクターの名前を決める」といったゲーム仕様の記述文がゲーム A とゲーム B の双方に含まれる場合、共通する要素が一点あるものとしてカウントした。

(5-2)自然言語処理技術によるサポートを活用した上での目視での類似性判断：

  ゲーム仕様の合計数の組み合わせは各 24 個のゲームで 40 個～100 個程度あった。各ゲームのゲーム仕様の延べ数を合計したところ 2175 個（重複を除外すると 1566 個）となり、ゲーム仕様それぞれの関係性のペアは 500 万個弱（2175 個の 2 乗）となってしまった。これは全てのゲーム仕様の表記の類似・一致をチェックする作業について人間が目視で実施できる数値ではない。そのため、ゲーム仕様の文として記述した内容に

---

[1] ゲームの内容確認作業としては、基本的には調査者全員が 2 時間以上ゲームを起動して、実際に触れることとした。2 時間のゲームプレイではゲーム仕様確認が不十分であると思われた点については、別途攻略本を参照するものとした。

[2] 具体的な作業としては、プログラムを用いて（python のコード）文の完全一致の判定を自動的におこなった。

ついて、まずゲーム仕様の記述文相互の意味的な類似性について、高度な自然言語処理のエンジンである BERT[3]を用いて類似性を測定し、近い意味を持つと思われるゲーム仕様の記述文を選出し、記述文の類似度が特に高いと判定された 3 件を選出した[4]。

　これにより、実際に人間が目視で類似性をチェックする作業が実際に可能な作業量まで縮小できたため、4698 ペア[5]のゲーム仕様の記述文の組み合わせについて実際にゲームの内容を確認した人間がその類似性を判定した。

　この作業を通じて、ゲーム仕様の記述文の文言が完全一致していなくとも、意味的な類似性が高いと判断されるゲーム仕様が 2 つのゲーム間に共通して含まれるケースについても「共通する要素をもつゲーム仕様をもつもの」として処理を行った。たとえば「メイン画面では 7 つのコマンドが選べる」という記述文と、「メニューでは 6 つのコマンドが選択できる」という記述文は、意味的な類似性が高いと判断[6]し、共通のゲーム仕様を持つ作品としてカウントしている。

　また、本手法の限界として、上記のような類似性の判定を行っているとはいえ、同じゲーム仕様でも、ゲームによって呼び方が異なる場合について、その際に同一のゲーム仕様として分析を行っていない場合がある。先述の通り、複数人がゲーム仕様の類似性について判断してはいるものの、人間が判断するにせよ、アルゴリズムによって意味の判定をするにせよ完全な判断ができるわけではない。本手法は中規模のビデオゲームの作品群の関係を分析するための現

---

[3] Google が開発した自然言語処理モデルであり、2018 年より一般に公開されている。

[4] より、詳細なプロセスとしてはまず BERT の判定において類似性が上位 12 位までとなるを選出した。その上で、類似性が高いと思われるゲーム仕様の文がその中に含まれていないかを人力で、100 の作品（100*12=1200 のゲームの記述文）を確認したところ、ほぼ類似と言えるゲーム仕様は概ね類似性が上位 3 位以内までに存在していることがわかった。そのため、BERT の類似性の判定において、類似性が上位 3 位以内までと判定されたものについて類似するゲーム仕様が含まれる可能性が高いため、BERT の判定による類似性が 3 位以内のものを一覧してチェックできる状態にした。

[5] 重複を除いた 1566 のゲーム仕様の記述文の類似度上位 3 件であるため、1566*3=4698 のペアとなる。一個のペアにつきおよそ 20 秒程度で判断すれば、一人あたり 26 時間でチェック作業が可能であると判断し、これを実施した。

[6] この点がもっとも属人的な判断が入り込んでしまう可能性が高い要素になるであろうと思われる。たとえば（1）「コマンドが 9 つ選択できる」と「コマンドが 2 つ選択できる」のペア（2）「バトル評価に応じて報酬が変化する」と「バトルの順位に応じて報酬を獲得できる」のペア　などの場合に類似性が高いかどうかの判断には、どうしても評価者による偏りが出てしまうため、複数人による目視を行っている。

時点で実施しうる最も妥当と思われる手続きとして実施されているものである。

上記の作業を得た上で次のような形式の作品間のゲーム仕様の共通度を示す表形式のデータを作成する。下記、表 3-1 に例として、その一部を示す[7]。

表 3-1：各ゲーム間の共通度を示すデータの一部（元データは、2175 行 28 列）

| GameMechanics | GameTitle | 卒業 Graduation | ときめきメモリアル | Jリーグプロサッカークラブをつくろう！ | 実況パワフルプロ野球3 |
|---|---|---|---|---|---|
| 育成コマンドには所持金を増加さ | プリンセスメーカー | 0 | 0 | 0 | 0 |
| 育成コマンドの中に休息コマンド | プリンセスメーカー | 0 | 0 | 0 | 0 |
| コスト系パラメータを消費しすぎ | プリンセスメーカー | 1 | 1 | 1 | 1 |
| 特定のパラメータが0になると状 | プリンセスメーカー | 1 | 1 | 1 | 0 |
| 年齢が上がると育成コマンドが増 | プリンセスメーカー | 0 | 0 | 0 | 0 |
| 時期イベントがある | プリンセスメーカー | 1 | 1 | 0 | 1 |
| 時期イベントでは育成対象キャラ | プリンセスメーカー | 1 | 1 | 0 | 0 |
| 時期イベントの結果が良いと普段 | プリンセスメーカー | 1 | 0 | 0 | 1 |
| 育成コマンドの1つにRPG風のシ | プリンセスメーカー | 0 | 0 | 0 | 0 |
| 育成期間終了時のパラメータによ | プリンセスメーカー | 1 | 1 | 0 | 0 |
| 育成コマンド内にさらに複数種類 | プリンセスメーカー | 1 | 0 | 1 | 1 |
| ゲームの最初にプレイヤーキャラ | 卒業 Graduation | 1 | 1 | 0 | 0 |
| ゲームの最初にプレイヤーキャラ | 卒業 Graduation | 1 | 1 | 0 | 0 |
| ゲームの最初にプレイヤーキャラ | 卒業 Graduation | 1 | 1 | 0 | 0 |
| ゲームの最初にプレイヤーキャラ | 卒業 Graduation | 1 | 0 | 0 | 0 |
| 血液型はプレイヤーキャラクター | 卒業 Graduation | 1 | 1 | 0 | 1 |

(6)作品間のゲーム仕様のネットワークを図示する

上記のような基礎データを整備した上で、Juul(2016)においてパズルゲームを例に示されている作品間ネットワークの表示のための手法を、より標準化した手法で示す。

・ 作品間の関係性の可視化：共通度を、ネットワークの線の太さによって示す（4.2 および、4.6）

・ クラスター分析によるゲーム間の関係性の近さの可視化（4.3）

---

[7] 元データとなる CSV については、下記 URL にてオープンデータとして公開している。
https://github.com/hiyokoya6/RisingSimulation/

- 当該ジャンルおよびサブジャンルにおける一般的な仕様の確認（4.4）

(7)具体的なゲーム仕様の事例提示

より具体的に当該ジャンルにおいてよく見られる仕様をとりあげて示すことで、当該ジャンルの傾向を示す。(4.5)

上記の作業を通じて、ジャンルの範囲および作品間の関係性を示すことで、調査者の属人性を可能な限り排除しつつ、データの再現性、検証可能性を高めたゲームジャンルの範囲の提示および系譜を示すことが期待できる。

## 4.事例研究：育成ゲームの系譜（2020 年まで）

### 4.1 育成ゲームの範囲の同定

本節では、「育成ゲーム」の概念の登場・普及時期を調査した後、ゲームプレイを含めたより詳細な調査を実施するゲームタイトルを選定する。

まず、育成ゲームがいつ頃から、どういったジャンルとして捉えられているのかを知るため、育成ゲームを説明する複数の雑誌記事や書籍、web サイトを選定した。その結果、最終的な調査対象となったのが表 4-1 である。

表 4-1：対象とした言説の一覧

これらの言説をもとに育成ゲームの概念の登場、普及時期について、時系列に沿ってまとめる。以降、【】で示す言説とその番号は、表 4-1 の「番号」と対応している。

### 4.1.1 育成ゲーム概念の登場

育成ゲームの源流にあたるものとされるのは、まず、商業用ビデオゲーム以前に、イギリスの数学者であったジョン・ホートン・コンウェイ (John Horton Conway)が考案したプログラムの『Conway's Game of Life』（1970）【言説 9, p.54】が最も古い。家庭用ゲームとしては、小人の世話をする『Little Computer People』（1985, アクティビジョン, Amiga 等）【言説 9】、子犬を育成する『パピー・ラブ』（1986, Tom Snyder Productions, Inc., マッキントッシュ）【言説 9】、『ベストプレープロ野球』（1988, アスキー, FC）【言説 17, p82】、飼育ゲームの『アップルタウン物語』（1987, DOG(スクウェア), FC）【言説 17, p83】、が初期のタイトルとして散見される。

1991 年には、女の子を育てるゲームとして、『プリンセスメーカー』（1991, ガイナックス, PC-98）が複数の記事で紹介される。この時点で、開発元であるガイナックスのチラシ【言説 3, p.104】の中では、『プリンセスメーカー』は「キャラクター育成 SLG」と表記されている。また、1991 年の段階で、国内では『信長の野望』（光栄）シリーズがすでに発売されていたが、海外より『SimCity』（1989, マクシス, PC）や『ポピュラス』（1989, エレクトロニック・アーツ, PC-9801）が輸入され、日本でも普及したことで、シミュレーションジャンルの拡大が起こったとされる【言説 2, p.98】。その中では、『プリンセスメーカー』は「子育てシミュレーションゲーム」【言説 2, p.104】と呼ばれ、子育てゲームという認識が強いものの一つのジャンルとしての認識はまだみられないことが確認できる。

### 4.1.2 育成ゲームのブーム

1993 年に入ると、育成ゲームのジャンルの認識が複数の言説から確認でき、中でもシミュレーションゲームから育成ゲームが切り離されて一つのジャンルとなっていることが歴史の整理ともに記述されている。【言説 6】では、「美少女育成シミュレーション」の歴史の特集が組まれており、図 4-1 の通り、シミュレーションゲームとは独立した形で育成ゲームが認識されている（p.7）。ただし、この時点ではメーカーへのアンケート調査から、育成シミュレーションゲームの定義は各社によって異なっていたという。そこで、改めてその定義を確認する試みが行われ、類似要素がある RPG などと比較して、「育成 SLG には敵という存在さえなく、純粋に育成するのを目的にしているのだ（p.9）。」と、あくまでもゲームの目的が戦闘ではなく、育成することにあることを強調する形で整理している。

図 4-1：SLG を中心とした最近のゲームの流れ（【言説 6, p.7】より引用）

1993 年時点では、育成ゲームが静かなブームを迎えているとし、より詳細に育成ゲームを分類、定義する試みが行われている。【言説 7】によれば、これまでの様々なビデオゲームのジャンルにも、例えばマリオ[8]がきのこでパワーアップする、RPG でのキャラクターの育成といった、育てる要素はありつつも、育成自体を目的としたゲームが現れた。そして、育成要素が

---

[8] マリオとは、『スーパーマリオブラザーズ』（1985, 任天堂, ファミリーコンピュータ）といったシリーズに登場するキャラクターである。

強いタイトルの対象は、図 4-2 の通り、都市、動物（馬、犬等）、人間（女の子）と分けられている。また、図 4-3 の通り、中でも人と動物を育成するゲームが「目的を持って育てる」に分類され、より近しいジャンルだと捉えていたものと思われる。

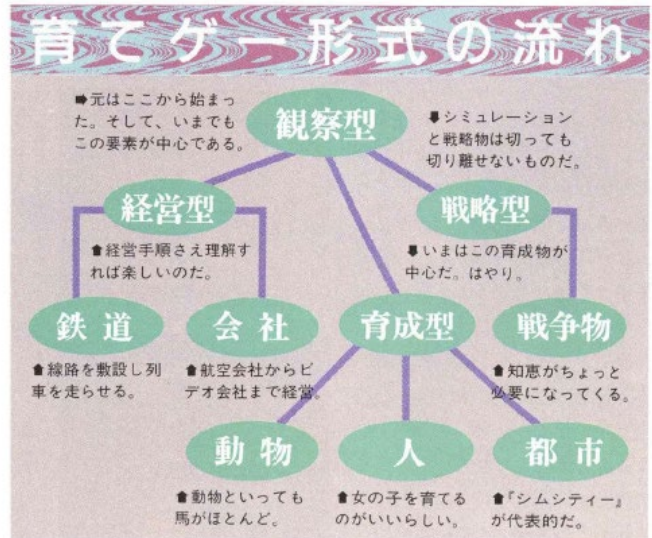

図 4-2：1993 年時点での育成ゲームの系譜（【言説 7, p.101】より引用）

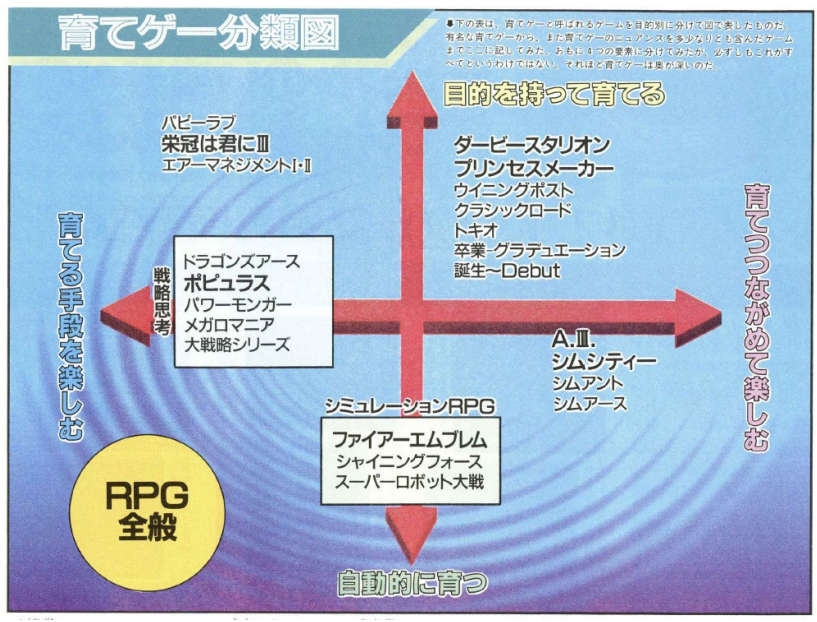

図 4-3：1993 年時点での育成ゲームの分類（【言説 7, p.101】より引用）

1994 年の【言説 8】では、「育成シミュレーション黄金時代の到来か？（p.112）」とその歴史を振り返る記事があり、特に 1991 年～1992 年は『プリンセスメーカー』と『卒業』の登場による育成型ギャルゲーの活性化、1993 年は PC ゲームから家庭用ゲーム機への移植と、『ときめきメモリアル』の人気による育成シミュレーション全盛期と記述がある。

### 4.1.3 育成ゲーム概念の定着

1995 年以降には、書籍中のエッセイとして育成ゲームの歴史を俯瞰する言説が複数発見された。その際には、大きく、人間、ペット（犬や猫等）、競馬、都市といった対象ごとに分けた整理がなされている【言説 9～20】。

この中でも『たまごっち』（1996, バンダイ, 携帯型電子ゲーム）の社会的なブームによって、電子ゲームにおいてペットを擬似的に育てることが注目を集めた。家庭用ゲームでも『がんばれ森川君 2 号』（1997, ソニー・コンピュータエンタテインメント, プレイステーション）といったタイトルが発売され、電子ペットをビデオゲームで育てるブームが起こる【言説 14, 15, 16】。さらに、『ポケットモンスター赤・緑』（1996, 任天堂, ゲームボーイ）は RPG ではあるが、キャラクターを育成し、それを交換することも可能であることから、育成要素も強い。

こういったタイトルは子どもに受け入れられ、育成ゲームは、大人のみならず子どもも含めた間口の広いジャンルになっていった。また、1999 年における育成ゲームは、「「労力を払って」「リアクションを得て」「モチベーションを維持し」「成長を見届ける」の 4 要素で構成されている」【言説 17, p.85】と説明されている。

### 4.1.4 現在の「育成ゲーム」

2000 年代以降は、ゲーム専門誌での特集はなくなり、これ以降は web サイトが主な情報源となる。ここでは引き続き著名シリーズの最新作が挙げられているほか、育成ゲームがスマー

トフォン等で手軽にプレイできるようになったこともあり、iOS/Android 等でのゲームタイトルが毎年複数リリースされていることがわかる【言説 21〜33】。

　現在、日本語版の、Wikipedia【言説 25】では、「育成ゲーム」は「シミュレーション」ゲームの下位ジャンルとされている。図 4-2 では、A.III と略されている『A列車で行こう』シリーズは、都市開発や企業経営が育成というよりも「経営」という語が一般に使用されるために、1990 年代後半には、育成シミュレーションではなく、経営シミュレーションであると認識されることもあった【言説 17, p.85】。同様に、現代においては、日本語版の Wikipedia では『A列車で行こう』シリーズや『Sim City』シリーズは「育成シミュレーション」【言説 25】ではなく、「経営シミュレーション」だと分類されている[9]。

　なお、英語との対応であるが、日本語版の Wikipedia における「育成シミュレーションゲーム」の項目を参照すると、対応する英語ページは"Life simulation game (Life Sim)"とされる。しかしながら、Life Simulation は、日本語における育成ゲーム概念の典型例をつくっていると推測される『プリンセスメーカー』が、必ずしも Life Simulation の概念の典型例を構築している様子が見受けられない。Life Simulation ジャンルは、複数のサブジャンルにまたがっており、サブジャンルとしての「Social Simulation[10]」の中に、『プリンセスメーカー』は位置づけられている。ただし、「Social Simulation」の概念範囲の中には、『どうぶつの森』シリーズや『Sims』なども含まれており、『プリンセスメーカー』と同様のゲーム仕様のセットを指す概念であるというよりは、ゲーム仕様ではない、ゲームの表象に関する主題を示す語彙であるように思われる。英語において、『プリンセスメーカー』のジャンル名は「Raising Sim」に分類されることも多い[11]が、こちらのほうがゲーム仕様をベースとする語彙として使われる側面が強いと思われる。

---

[9] Wikipedia 日本語版.「経営シミュレーションゲーム」2023 年 12 月 8 日 09:05最終更新版
https://ja.wikipedia.org/wiki/%E7%B5%8C%E5%96%B6%E3%82%B7%E3%83%9F%E3%83%A5%E3%83%AC%E3%83%BC%E3%82%B7%E3%83%A7%E3%83%B3%E3%82%B2%E3%83%BC%E3%83%A0 , 2023 年 12 月 12 日閲覧.
[10] Wikipedia. Social Simulation. 2023 年 11 月 18 日 14:34 (UTC)最終更新版
https://en.wikipedia.org/wiki/Social_simulation_game , 2023 年 12 月 20 日閲覧.
[11] TV Tropes. Raising Sim. https://tvtropes.org/pmwiki/pmwiki.php/Main/RaisingSim , 2023 年 12 月 20 日閲覧.

このように、「育成ゲーム」概念は国内外で異なっており、地域差が生じている。また、Brigham et al. (2021)による多言語によるビデオゲームジャンルの分類体系でも、"breeding"が対応する訳語とされるなど、言語間での語彙の範囲の対応が揺れている状況にある。

### 4.1.5 NDL Ngram Viewer から見た育成ゲームの語の普及

NDL Ngram Viewer のデータから、ビデオゲームにおける育成ゲームのジャンル普及タイミングを可視化したデータを下記に記載する。NDL Ngram Viewer は、国立国会図書館が提供するデジタル化資料の OCR テキスト化事業の成果物である全文テキストを活用した実験サービスで、OCR によって作成されたテキストデータから、出版年代ごとの出現頻度を可視化することが可能である [12]。このサービスでは現在のところ、主要なゲーム専門誌がデジタル化資料に含まれておらず [13]、また、図書については刊行年代が 1960 年代まで、雑誌については刊行年代が 1990 年代までの資料が主な対象となっている。そのため参考とはなるが、一般向けの週刊誌やマンガ雑誌、パソコン雑誌等からヒットした結果を見ていく。

前項での調査により、日本語圏において、「育成ゲーム」という概念が定着する以前のものとして、「育てゲー」「育成SLG（育成シミュレーション）」という言葉が 90 年代前半には、ほぼ同様の意味で使われていることがわかる。そこで、この3つの語を NDL Ngram Viewer にてテキスト検索を行った結果が、図 4-4 である。

---

[12] NDL Ngram Viewer. 国立国会図書館. https://lab.ndl.go.jp/ngramviewer/，（Accessed 2023-12-12）
[13] 以下サイトにおいて、デジタル化候補の雑誌がリスト化されている。国立国会図書館. https://www.ndl.go.jp/jp/preservation/digitization/periodicals.html，（Accessed 2023-12-12）．

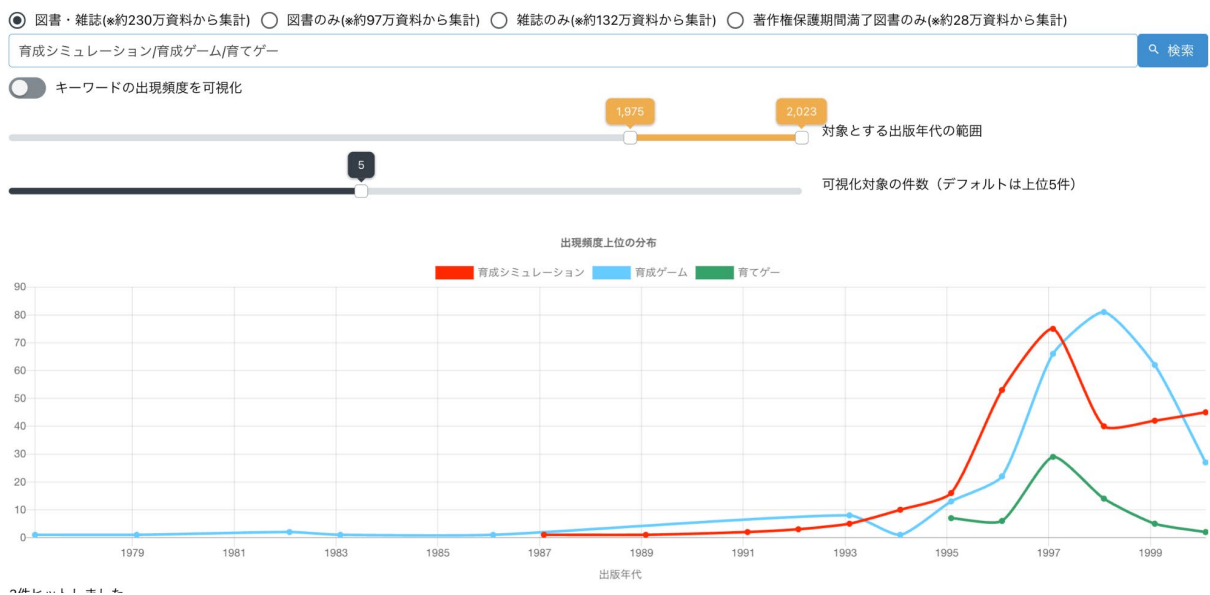

図 4-4：「育成ゲーム」を示す語の普及時期

このように、育成ゲームの概念の登場が一般的に見られるようになったのが 1980 年代後半、そして、急速に普及したのが 1995 年以降であったことがわかる。特に 1996 年の急増は、『たまごっち』の発売が大きく影響していると思われる。

### 4.1.6 育成ゲーム概念の整理

以上を整理し、まとめる。1993 年に登場した『プリンセスメーカー』を中心に、育成ゲームの隆盛が各雑誌記事に反映され、初めはシミュレーションゲームの一部として紹介されていた育成ゲームは、単独でのジャンルとしてその歴史も関連記事で確認でき、一つのジャンルとしての確立をみることができた。ゲームの目的は戦闘ではなく育成することにあること、育成の対象が人間、犬や猫等のペット、競走馬に大きく分かれている。

1995 年ごろに、ゲーム専門誌における育成ゲームの記事特集は減少し、代わりにインタビュー記事やジャンル別の外観を論じる記事、また、育成ゲームを包括的に特集する雑誌や図書において、育成ゲームに関する記事やエッセイが見受けられた。しかし、1990 年代後半の電子ペ

ットブームを最後に、2000年以降になると著名なシリーズやペットの育成ゲームが引き続き多数発売されているにもかかわらず、雑誌や書籍においては育成ゲームの特集記事がほとんど存在しなくなる。

近年では、webサイトにおいてモバイル端末やブラウザゲームの著名なタイトルがいくつか確認できたが、記事がアップデートされて残っていないため、2000年〜2020年ごろは記事自体が見つからない空白の期間となっている。

### 4.1.7 調査対象とする育成ゲームの選定

表4-1中の複数の言説から育成ゲームとして紹介されているゲームタイトルを抽出し、4.2節以降、それらのゲームプレイを通じたより詳細な検討の対象とすることにした。

まず、表4-2は、表4-1で示した言説で取り上げられているゲームタイトルを計上し、2つ以上の参考文献でみられたゲームタイトル等をまとめたものである。

表 4-2：育成ゲームとして複数回とりあげられたゲームタイトル

| タイトル | 言及回数 |
|---|---|
| 『プリンセスメーカー』（1991, ガイナックス, PC-98他） | 15 |
| 『ダービースタリオン』（1991, アスキー, ファミリーコンピュータ/スーパーファミコン/PlayStation/セガサターン） | 9 |
| 『卒業 ～Graduation～』（1992, ジャパンホームビデオ, PC-98/3DO/セガサターン/Windows） | 8 |
| 『たまごっち』（1996, バンダイ, 携帯型ゲーム） | 7 |
| 『モンスターファーム』（1997, テクモ, PlayStation） | 7 |
| 『SimCity』（1989, マクシス, Commodore 64他） | 6 |
| 『ウマ娘 プリティーダービー』（2021, Cygames, iOS / Android） | 5 |
| 『ウイニングポスト』（1993, 光栄, PC-98） | 4 |
| 『A列車で行こう3』（1990, アートディング, PC-98他） | 4 |
| 『なめこ栽培キットDeluxe 極』（2020, BEEWORKS GAMES, iOS / Android） | 4 |
| 『アイドルマスター シャイニーカラーズ』（2018, BANDAI NAMCO Entertainment, iOS / Android） | 4 |
| 『モンスターファーム2』（1999, テクモ, PlayStation） | 4 |
| 『プリンセスメーカー2』（1993, ガイナックス, PC/DOS/FM-T/PCエンジン/セガサターン/3DO ） | 3 |
| 『SimEarth』（1990, マクシス, PC） | 3 |
| 『がんばれ森川君2号』（1997, ソニー・コンピュータエンタテインメント, PlayStation） | 3 |
| 『マイたまごっち』（2021, BANDAI NAMCO Entertainment Europe, iOS / Android） | 3 |
| 『パピーラブ』(1986, Tom Snyder Productions, Inc., マッキントッシュ他) | 3 |
| 『デジタルモンスター』（1997, バンダイ, 携帯型電子ゲーム） | 3 |
| 『Jリーグ プロサッカークラブをつくろう！2』（1997, セガ・エンタープライゼス, セガサターン） | 3 |
| 『アイドルマスター　ステラステージ』（2017, バンダイナムコエンタテインメント, PlayStation4） | 3 |
| 『Birthdays the Beginning』（2017, アークシステムワークス, PlayStation4） | 3 |
| 『Winning Post 9 2021』（2021, コーエーテクモゲームス, Nintendo Switch） | 3 |
| 『実況パワフルプロ野球』（1994, コナミ, スーパーファミコン） | 2 |
| 『FOXY』（1990, エルフ, PC-88） | 2 |
| 『ストロベリー大作戦』（1990, フェアリーテール, PC-98） | 2 |
| 『誕生-Debut-』（1993, NECアベニュー, PC/PCエンジン/3DO/セガサターン/Windows） | 2 |
| 『ときめきメモリアル』（1994, コナミ, PCエンジン/PlayStation/スーパーファミコン/セガサターン他） | 2 |
| 『子育てクイズマイエンジェル』（1996, ナムコ, アーケード）（1997, ナムコ, PlayStation） | 2 |
| 『クラシックロード』（1993, ビクターエンターテイメント, スーパーファミコン） | 2 |
| 『トキオ』（1995, アートディンク, PC-98他） | 2 |
| 『ベストプレープロ野球』（1988, アスキー, ファミリーコンピュータ） | 2 |
| 『リトル・コンピュータ・ピープル』（1985, アクティビジョン, AMIGA他） | 2 |
| 『A列車で行こう』(1985, アートディンク, FM-7他) | 2 |
| 『はねろ！コイキング』（2017, SELECT BUTTON, iOS / Android） | 2 |
| 『プロサッカークラブをつくろう! ロード・トゥ・ワールド』（2018, SEGA, iOS / Android） | 2 |
| 『艦隊これくしょん -艦これ-』（2013, DMM.com, ウェブブラウザ） | 2 |
| 『BOXER'S ROAD』（1995, NEW, PlayStation） | 2 |
| 『アップルタウン物語』（1987, DOG(スクウェア), ファミリーコンピュータ） | 2 |
| 『パワーモンガー』(1990, エレクトロニック・アーツ, AMIGA) | 2 |
| 『モンスターファーム2』（2020, コーエーテクモゲームス , iOS / Android） | 2 |
| 『ダービースタリオンⅡ』（1994, アスキー, スーパーファミコン） | 2 |
| 『ポケットモンスター赤・緑』（1996, 任天堂株式会社, ゲームボーイ） | 2 |
| 『シムシティ2000』(1994, マクシス, Windows他) | 2 |
| 『ウルトラ怪獣モンスターファーム』（2022,バンダイナムコエンターテインメント,Nintendo Switch） | 2 |
| 『天穂のサクナヒメ』（2020, マーベラス, PlayStation4 / Nintendo Switch） | 2 |
| 『じゃんがりあん物語』（2021, サクセス, Nintendo Switch） | 2 |
| 『ダービースタリオン』（2020, ゲームアディクト, Nintendo Switch） | 2 |
| 『放置少女～百花繚乱の萌姫たち～』（2017, C4games, iOS / Android / ウェブブラウザ） | 2 |
| 『刀剣乱舞 ONLINE』（2015、EXNOA, ウェブブラウザ） | 2 |
| 『ようとん場MIX』(2017,JOE,iOS/Android) | 2 |
| 『ペンギンの島』(2019,HABBY,iOS/Android) | 2 |
| 『元祖 なめこ栽培キット』(2021,ビーワークス,iOS/Android) | 2 |

前項までと、表 4-2 を元に、育成ゲームに特徴的なゲーム仕様を抽出するために選定した育成ゲームを、表 4-3 の通り 25 タイトル決定した。選定の基準を以下に述べる。

　全体を通して、ゲームタイトルは、発売後、事後的に評価が高まることもあること、資料数が限定されると考えられることから、選定の期間は 2020 年までとし、満遍なく幅広い年代から選定するよう留意した。その上で、選定に際しては 3 群を設定した。まず、1 群目は、表 4-2 で示した通り、複数の言説において育成ゲームであると記述されたゲームタイトル、あるいはシリーズのタイトルである。

　次に、2 群は、表 4-2 での言及回数は 1〜2 回ではあるが、売上や普及度、シリーズ中において特徴的なタイトル等から選択した。特に、日本における育成ゲームは、人間（女の子）や競馬を対象とするタイトル数が相当数販売されたことを受け、『エターナルメロディ』、『悠久幻想曲』、『ウイニングポストワールド』といったタイトルを選択した[14]。

　最後に、3 群は、育成ゲームであるとはジャンル分けされる傾向はないが、育成要素を持つもので、売上数が多かったものである。具体的には、主人公や仲間キャラクターを育成していく RPG 等であり、『ポケットモンスター ルビー・サファイア』や、『ドラゴンクエスト IX 星空の守り人』、『ドラゴンクエストモンスターズジョーカー 2』である。また、2000 年以降は育成ゲームを特集する言説の掲載が見られなくなったものの、動物を飼育するゲームタイトルは引き続き多く発売されており、その中でも特に売上の多かった『nintendogs 柴&フレンズ』を選定した。

　なお、ゲームタイトルによっては、初出タイトルの入手やプレイすることが困難であったものがある。『プリンセスメーカー』は 1991 年 5 月 24 日（PC-98）、『卒業〜Graduation〜』は 1992 年 6 月 25 日（PC-98）が初出タイトルであるが、入手が困難であったため、初出に近

---

[14] ウイニングポストについてはシリーズとして言及されているケースが多く特に作品を限定しているものが少なかったため、調査対象としての選定方法について研究チーム内で協議のうえ、ゲーム仕様をリストアップ対象としては『ウイニングポストワールド』のみを用いることとし、ただし一部具体的な個別のゲーム仕様の提示（4.5）においては、適宜シリーズ作品として『ウイニングポスト 2』を例示に含めるものとした。

い同一タイトルをそれぞれ選択した。また、iOSやブラウザでのゲームは、サービスが終了していたり、アップデートによってリリース当時と現在とでは仕様に大幅な変更が生じることも多数ある。そのため、できる限り初出日に近い日時に動画投稿サイトにて実況動画としてアップロードされている映像を見ることで調査対象とした。

表 4-3：選定したゲームタイトルの一覧

| 番号 | ゲームタイトル | 発売日 | 開発/発売者 | ハード |
|---|---|---|---|---|
| 1 | ベスト競馬・ダービースタリオン（ダビスタ） | 1991年12月21日 | アスキー | ファミリーコンピュータ |
| 2 | 卒業 ～Graduation～（卒業） | 1993年7月30日 | NECアベニュー | PCエンジン |
| 3 | ときめきメモリアル（ときメモ） | 1994年5月27日 | コナミ | PCエンジン |
| 4 | プリンセスメーカー（プリメ） | 1995年1月3日 | 日本電気ホームエレクトロニクス | PCエンジン |
| 5 | ウイニングポスト2（ウイポス2） | 1995年3月18日 | 光栄 | スーパーファミコン |
| 6 | Jリーグプロサッカークラブをつくろう!（サカつく） | 1996年2月23日 | セガ・エンタープライゼス | セガサターン |
| 7 | 実況パワフルプロ野球3（パワプロ3） | 1996年2月29日 | コナミ | スーパーファミコン |
| 8 | エターナルメロディ（エタメロ） | 1996年11月22日 | メディアワークス | プレイステーション |
| 9 | ゲームで発見!!たまごっち（たまごっち） | 1997年6月27日 | バンダイ | ゲームボーイ |
| 10 | 悠久幻想曲（悠久） | 1997年7月18日 | メディアワークス | セガサターン |
| 11 | モンスターファーム（モンファ） | 1997年7月24日 | テクモ | プレイステーション |
| 12 | 子育てクイズ マイエンジェル（子育て） | 1997年11月13日 | ナムコ | プレイステーション |
| 13 | プリンセスメーカー ～ゆめみる妖精～（プリメ・ゆ） | 1998年6月18日 | ガイナックス | セガサターン |
| 14 | ポケットモンスター ルビー・サファイア（ポケモン ルビサファ） | 2002年11月21日 | ゲームフリーク/任天堂 | ゲームボーイアドバンス |
| 15 | nintendogs 柴&フレンズ（nintendogs） | 2005年4月21日 | 任天堂 | ニンテンドーDS |
| 16 | アイドルマスター（アイマス） | 2007年1月25日 | バンダイナムコゲームス | Xbox360 |
| 17 | ウイニングポストワールド（ウイポスワールド） | 2009年4月2日 | コーエー | プレイステーション3 |
| 18 | ドラゴンクエストIX 星空の守り人（ドラクエ9） | 2009年7月11日 | スクウェア・エニックス | ニンテンドーDS |
| 19 | ドラゴンクエストモンスターズ ジョーカー2（ジョーカー2） | 2010年4月28日 | スクウェア・エニックス | ニンテンドーDS |
| 20 | おさわり探偵 なめこ栽培キット（なめこ） | 2011年6月30日 | ビーワークス | ios |
| 21 | パズル＆ドラゴンズ（パズドラ） | 2012年2月20日 | ガンホー・オンライン・エンターテイメント | ios |
| 22 | 艦隊これくしょん -艦これ-（艦これ） | 2013年4月23日 | 角川ゲームス | ブラウザ |
| 23 | 実況パワフルプロ野球（パワプロアプリ） | 2014年12月18日 | コナミ | ios |
| 24 | 刀剣乱舞（とうらぶ） | 2015年1月14日 | CLARITY STUDIO/EXNOA | ブラウザ |
| 25 | Fate/Grand Order（FGO） | 2015年8月12日 | DELiGHTWORKS/アニプレックス | ios |

また、各タイトルにおいて、詳細な仕様の確認が難しい箇所については、適宜、攻略本を用いて仕様を確認した。参照した攻略本は下記の通りである。

- Sega Saturn Magazine 編集部, アミューズメント書籍編集部編. (1997). 悠久幻想曲公式攻略ガイド. 電撃攻略王. メディアワークス.
- アスペクト編集部. (1997a). ゲームで発見!! たまごっち 公式ガイドブック. アスペクト.
- アスペクト編集部. (1997b). モンスターファームブリーダーズガイド. アスペクト.
- アスペクト編集部. (1998). 子育てクイズマイエンジェル オフィシャルガイドブック. アスペクト.
- サラブレッド探偵局. (1995). ウイニングポスト 2 パーフェクトブック. 光栄
- スクウェア・エニックス（監修）. (2009). ドラゴンクエストIX星空の守り人公式ガイドブック 秘伝最終編. 株式会社スクウェア・エニックス.
- スタジオベントスタッフ. (2010). ドラゴンクエストモンスターズジョーカー 2 公式ガイドブック. 株式会社スクウェア・エニックス.
- 電撃 PlayStation＆電撃 SEGAEX 編集部. (1996). エターナルメロディ公式攻略ガイド. メディアワークス.
- 電撃 PlayStation 編集部＆電撃 SEGASATURN 編集部. (1997). 悠久幻想曲公式攻略ガイド. メディアワークス.
- ファミ通書籍編集部. (2005). ニンテンドッグスと暮らす本. エンターブレイン.
- 編集部. (2009). ウイニングポストワールド マスターガイド. 株式会社 光栄.
- 元宮秀介＆ワンナップ編著. (2002). ポケットモンスター ルビー・サファイア 公式ぼうけんクリアガイド. 株式会社メディアファクトリー.

なお、表 4-3 中の「ゲームタイトル」には、たとえば「ベスト競馬・ダービースタリオン（ダビスタ）」のように記入されているが、これは「正式タイトル（略称）」という意味である。これ以降において、一部の作品名称について、図示する際に表示上の都合等から一部略称を用いる。略称については、一般にゲーム業界、ゲームファンの間で流通している略称に準拠している。

## 4.2 「育成ゲーム」の系譜

まず、1566個のゲーム仕様のうち、それぞれのゲーム仕様の共通度合いを可視化する。下記の図4-5は、それぞれのメカニズムが、先発・後発のゲームとどれだけ共通しているかを示すものである。

ゲームとゲームをつなぐ線の太さは、ゲーム相互に共通するゲーム仕様の数を示している。

図4-5：各ゲームの相互のゲーム仕様の共通度合い（サンキー図による出力を利用 [15]）

---

[15] サンキー図とは、ある状態から別の状態へ、またはある時点から別の状態への流れ/移動/変化を強調するデータ視覚化手法である。複雑な多段階プロセスを提示するための手法である。同一リソースの流

厳密にどのゲームがどのゲームに影響を受けたのかについては、開発者のインタビュー等に依拠しなければ明らかにならないが、それぞれのゲーム仕様がどの程度まで共通するものであるかは上記の図によって示すことができる。

　また、ゲームの縦幅の太さは共通する仕様の数の延べ数の最大値を示しており、共通するゲーム仕様が多いゲームであるほど縦幅の長さが太くなっている。概してゲームの仕様の総量が少ないゲーム（『おさわり探偵なめこ栽培キット』など）では縦幅が短い。また、ゲームの仕様の総量が多いか、もしくは、今回の調査対象となった作品と共通する点を数多くもつ作品は縦幅が長くなる傾向がある。さまざまな要素を実装している『ポケットモンスター　ルビー・サファイア』『ドラゴンクエストジョーカー2』『刀剣乱舞』『Fate/Grand Order』といったタイトルは縦幅が長くなっている。

　図4-5から見て明らかな通り、今回の調査対象となった作品は相互に共通する仕様を数多くもっている。ほぼ全ての作品は、既存作品の要素を数多く継承しながら新しいゲームを構築してきたことが伺える。

## 4.3 育成ゲームに特徴的なゲーム仕様

### 4.3.1 調査対象ゲーム群におけるゲーム仕様をもとにした内容の類似性

　では、具体的にそれぞれの作品がどの程度類似する内容となっているのかを確認していきたい。上記の4-5のネットワーク図（サンキー図）を精緻に観察すれば、各ゲームの類似度を確認することは不可能ではないが、一見してわかりやすい可視化になっているとは言えない。そのため、ゲーム仕様の共通度のデータをもとにクラスター分析を実施した。図4-6として、最遠隣法と、ウォード法によりゲーム仕様の類似度を樹形図にしたものを示す。

---

入・流出というという形で使用される図であるが、図4-5ではあるゲーム仕様の共通度合いを示すために使用している。

この内容分類の樹形図は、4.1 で示した言説群もしくは、実際のゲームプレイを通じた調査者らの印象評価、およびゲームの開発関係者らの証言等のいずれかにおいて概ね整合的に解釈できる内容となっている。

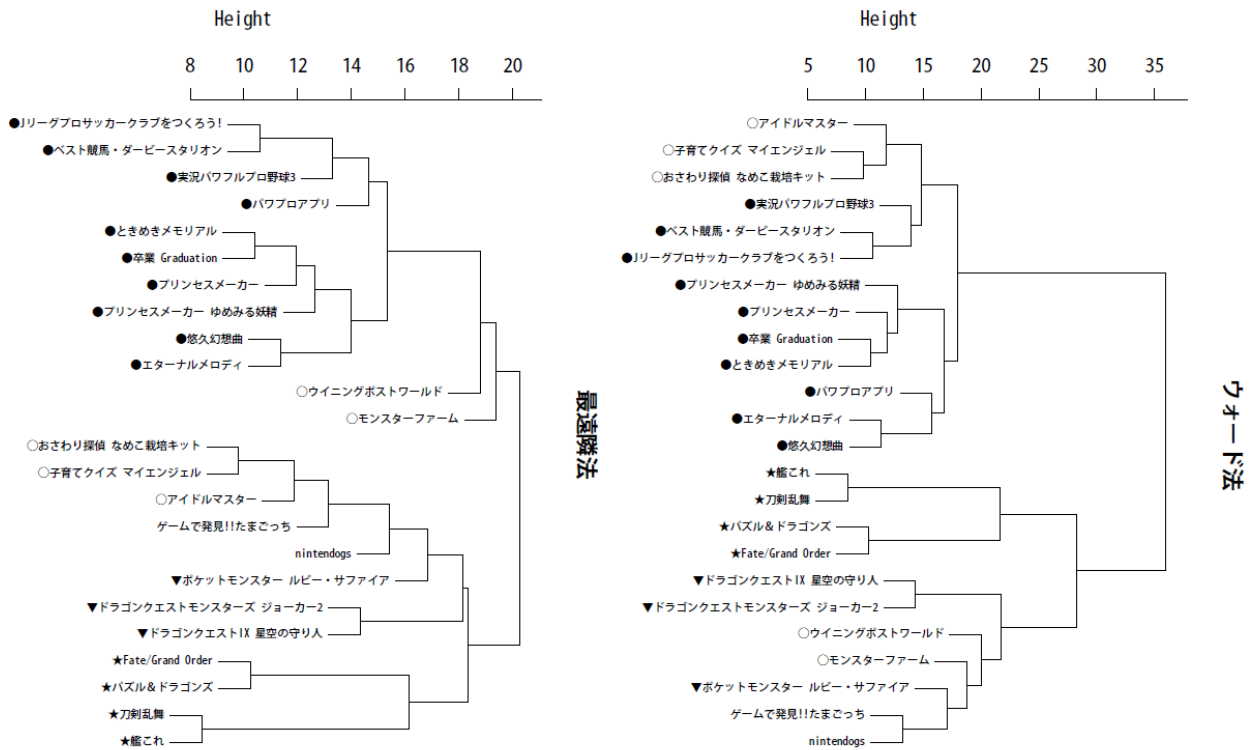

図 4-6：最遠隣法とウォード法によるゲーム仕様の類似度（樹形図）

上記の観点から、図 4-6 の 2 つの樹形図から解釈可能と思われる作品群を 5 群に分けて示す。

## （1）「育成ゲーム」の典型作品群（冒頭●マーク, 樹形図上部）

もっとも大きな 2 分類は概ね「育成ゲーム」として挙げられやすい作品群とそれ以外の作品群を分けていると解釈できる。図 4-6 中のそれぞれの上部の作品群はいずれも 4.1 で育成ゲームとして多くの言説で取り上げられたもの（第一群）が含まれており、もう半分はそれ以外の

作品が含まれる傾向が強い。特に上部の群は「育成ゲーム」概念と概ね一致した群とみなすことができるだろう。

　『ベスト競馬・ダービースタリオン』、『Jリーグプロサッカークラブをつくろう!』、『実況パワフルプロ野球3』、『プリンセスメーカー』、『卒業〜Graduation〜』、『ときめきメモリアル』、『エターナルメロディ』、『悠久幻想曲』、『プリンセスメーカー 〜ゆめみる妖精〜』、『パワプロアプリ』の10作品はいずれのクラスタリング手法においても、育成ゲームの特徴的な作品として検出できている。この10作品が育成ゲームの作品群に属することは頑健な結果であると見做すことができるだろう。開発者インタビューなどからも、明確に相互の影響関係が明らかになっている作品[16]も含まれており、内容的な妥当性は高いものと言える。

　『エターナルメロディ』や『悠久幻想曲』は、育成ゲームとして言及されている回数は、いずれも一回でありそこまで多くないが、調査者らの実感としては90年代中盤の育成ゲームの典型性に近い内容であり、これらは雑誌等の言説のみでは明らかにできなかったゲームの内容水準であるが、このクラスター分析によって近似性を示すことができている。

　また、下半分の作品群と比べた時、クラスターの分割が生じている水準（height）が低くなっており、これは上半分の群がいずれもゲーム仕様における類似性が高いことを示している。

（２）「育成ゲーム」の準典型作品群（冒頭○マーク）

　『子育てクイズマイエンジェル』、『アイドルマスター』、『おさわり探偵 なめこ栽培キット』、『ウイニングポストワールド』、『モンスターファーム』の5作品は、ウォード法か最遠隣法かのクラスタリング手法によって「育成ゲーム」を中心とした群の中に含まれるかどう

---

[16] たとえば、『パワプロ』シリーズのサクセスモードが『ときメモ』を直接に参考にして生まれていることは、『パワプロ』シリーズの開発者である山口剛自身がインタビューにおいて次のように述べている。「『サクセスモード』は、実は他のゲームを参考にして生まれたんです。弊社の一大コンテンツの『ときめきメモリアル』なんです。」（10th full count.2019「「パワプロ」「プロスピ」制作秘話「サクセスモード」は「ときメモ」から誕生!?」https://full-count.jp/2019/10/15/post573767/3/ ＜2023年12月19日閲覧＞

かが変化している。これらの作品は「育成ゲーム」の典型例と比較した場合に、相対的に典型的な特徴を備えていない側面が見られる作品群あると言える。

育成ゲームをめぐる言説においてしばしば取り上げられる作品であるが、これらは「育成ゲーム」としてくくられる作品の中でも、『プリンセスメーカー』の系譜にある育成 SLG などとは、ゲーム仕様に異なる点の多い作品群である。

言説として「育成ゲーム」としてくくられやすいがゲーム仕様がある程度まで異なる群を適切に別のクラスターとして区別できているものと解釈できるだろう。

（３）育成要素のあるソーシャルゲーム群（冒頭★マーク）

典型的な育成ゲームではないと思われる群に含まれる作品のうち『艦これ』、『刀剣乱舞』、『パズル＆ドラゴンズ』、『Fate/Grand Order』は、言説単位で見た場合、育成ゲームとされる回数はそこまで高頻度ではない（２回〜０回）作品群である。これらの作品は 2010 年代にいずれもスマートフォンで遊ぶことのできるソーシャルゲームとして人気を呼んだ作品である。いずれもオンラインのサーバーを経由して遊ぶものであり、育成要素が含まれ、抽選の仕様（ガチャ）、行動回数の制限値があるといった特徴が類似している。

また『刀剣乱舞』と『艦これ』はゲーム仕様が極めて近似したものであることを開発関係者自身らが語っているものでもあり[17]、この２作が同一のクラスターになっているのは、本データの妥当性を示すものであると言えるだろう。

（４）育成要素のあるロールプレイングゲーム（RPG）群（冒頭▼マーク）

---

[17] 『艦これ』開発者の岡宮道生がインタビューにおいて次のように述べている。「テストプレイさせてもらったら基本のシステムとかもそのまんまで。その時に初めてそのゲームの企画概要書を見せてもらったんですが、『艦これ』の画面がそのまま載ってたり、仕様に関しての説明も「屯所　※基本『艦これ』と同じ」「『艦これ』でいう○○こと××」とか書いてあるだけだったりして。…（略）…その後、担当のプロデューサーに「あれ、まんまコピーでしょ？」って聞いたら「すいません」って謝られましたが（笑）。」 "「戦闘シーンが無茶苦茶カッコいい」『艦これ』の産みの親がアニメを語る". DMM ニュース (2015 年 1 月 31 日). https://dailynewsonline.jp/article/912828/, 2023 年 12 月 17 日閲覧.

典型的な育成ゲームではないと思われる群に含まれる作品のうち、『ドラゴンクエスト IX』、『ドラゴンクエストモンスターズジョーカー 2』、『ポケットモンスター ルビー・サファイア』は、今回育成ゲームとみなされるもの以外のうち育成要素をもつ有名作品として選定した RPG の有名作品群であるが、これらについても適切に下半分のクラスターの中に含まれていることが伺える。

（5）育成ゲーム境界例群（冒頭マークなし）

『nintendogs』、『ゲームで発見!!『たまごっち』については、いずれの手法においても育成ゲームの典型とされる群に含まれていない。内容的にも『プリンセスメーカー』等のゲーム仕様とは独立性が高いことが上記のデータから伺える。

### 4.3.2 育成ゲームにおけるゲーム仕様のクラスター

同様にゲーム仕様についてのクラスター分析を図 4-7 に示す。ただし、分析の対象となったゲーム仕様全体の数は多岐にわたる。そのため、図 4-7 ではクラスター分析の対象は 8 個以上のゲームにおいて見られた要素のみの 65 を対象に取り扱っている。

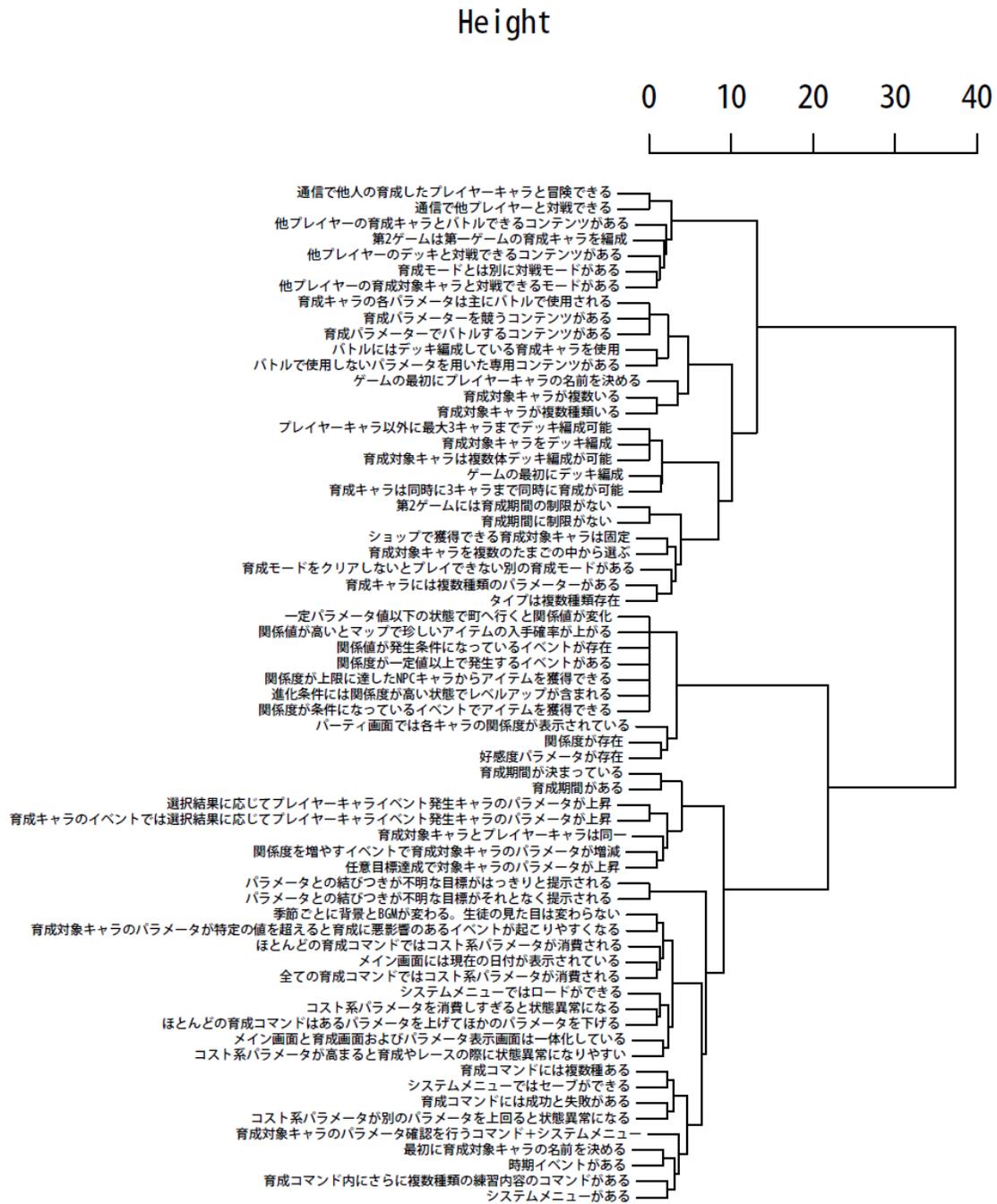

図 4-7：ゲーム仕様の樹形図（8 個作品以上で確認できたもの）

図 4-7 の樹形図に示したゲーム仕様は、この状態でも数が多く、関連が複雑なため作品間の関係性ほどに解釈が容易ではないが、概ね次のようなゲーム仕様がクラスターをなしているということが言えるだろう。

- 育成コマンドに関わる仕様（育成コマンドが複数あることを示すもの、育成のモードと育成したキャラを活用できるモードに関するもの）
- キャラクターとの関係値および、イベントに関わるもの
- キャラクターの育成に関わる仕様（回数制限の消費および回復に関するものなど）
- メニュー画面や UI に関わるシステムのロード／セーブ、日付表示などほとんどのゲームに共通とみられる仕様
- その他の育成関連の仕様
- その他の対戦要素に関わる仕様

## 4.4　当該ジャンルに一般的なゲーム仕様の分析

### 4.4.1 典型作品群に一般的なゲーム仕様

4.3.1 で示した「育成ゲーム」の典型作品群において頻繁にみられたゲーム仕様[18]は、表 4-4 の通りとなった。

なお、「カテゴリー」の括りは 4.3.2 に見られた傾向をもとに分析者によって判断したカテゴリーである。

表 4-4：「育成ゲーム」に頻繁にみられたゲーム仕様の一覧

---

[18] ここでは表示上の都合から、典型育成ゲーム群の 3 割以上（10 作品中 3 回以上）見られたものを示した。2 回以上の仕様については、別途公開しているデータを確認されたい。

| カテゴリー | ゲーム仕様の一覧 |
|---|---|
| 育成コマンドに関わる仕様 | ・ 育成コマンドがある<br>・ 育成コマンドには成功と失敗がある<br>・ 育成コマンド内にさらに複数種類の練習内容のコマンドがある<br>・ 全ての育成コマンドではコスト系パラメータが消費される<br>・ ほとんどの育成コマンドはあるパラメータを上げてほかのパラメータを下げる |
| 関係値に関わる仕様 | ・ 関係度が数値以外の方法で表示される<br>・ 関係度が存在する<br>・ 関係度の効果はイベントの発生に影響する<br>・ 関係度の効果はエンディングの変化に影響する |
| キャラクターに関わる仕様 | ・ 育成対象キャラクターが複数いる<br>・ 育成対象キャラクターとプレイヤーキャラクターは同一<br>・ 時期イベントでは育成対象キャラクターのパラメータによって結果が変わる<br>・ 時期イベントがある育成対象キャラクターにはボイスがついている |
| メニュー画面やUIに関わる仕様 | ・ メイン画面と育成画面は一体化している<br>・ メイン画面には日付が表示されている<br>・ システムメニューがある<br>・ システムメニューではロードができる<br>・ システムメニューではセーブができる<br>・ メイン画面には所持金が表示されている<br>・ ゲームの最初にプレイヤーキャラクターの名前を決める<br>・ ゲームの最初に育成対象キャラクターの名前を決める |
| その他の育成関連仕様 | ・ 育成期間が決まっている<br>・ コスト系パラメータを消費しすぎると状態異常になる<br>・ パラメータとの結びつきが不明な目標がそれとなく提示される |

### 4.4.2 準典型作品群

　同じく、4.3.1（2）で示した例となる作品群において頻繁にみられたゲーム仕様[19]を表4-5に示す。

---

[19] ここでは、4割以上（5作品中の2作品以上）のものを示した。

表 4-5：準典型作品群において頻繁にみられたゲーム仕様

| カテゴリー | ゲーム仕様の一覧 |
|---|---|
| 育成コマンドに関わる仕様 | ・ 育成コマンドが複数種ある<br>・ 育成コマンド内にさらに複数種類の練習内容のコマンドがある<br>・ 育成コマンド実行で日付が進行<br>・ 1ターンに1育成コマンドずつ実行 |
| 関係値に関わる仕様 | ・ 関係度は数値化されている<br>・ 関係度が存在する |
| キャラクターに関わる仕様 | ・ 育成対象キャラクターが複数いる<br>・ 育成対象キャラクターにはタイプがある |
| メニュー画面やUIに関わる仕様 | ・ ゲームの最初にプレイヤーキャラクターの名前を決める<br>・ ゲームの最初に育成対象キャラクターの名前を決める<br>・ システムメニューからセーブができる |
| その他育成関連仕様 | ・ 育成期間が決まっている |

　準典型作品群においても、共通する仕様として複数回登場する仕様は典型的な育成ゲーム群と概ね共通するものであることが伺える。ただし、個別の作品ごとの仕様のバラつきが典型作品群よりもやや大きくなっている傾向が伺える。

### 4.4.3 ソーシャルゲーム群

　同じく、4.3.1（3）で示したソーシャルゲームの作品群において頻繁にみられたゲーム仕様[20]を表 4-6 に示す。

---

[20] ここでは、6 割以上（5 作品中 3 作品以上）を示した。4.4.2 と合わせて 4 割以上で統一したいところであるが、準典型作品群は共通する仕様が比較的少ない。一方で、ソーシャルゲーム群は、互いに似たようなゲーム仕様を持つ側面が高く、4 割以上の作品の基準で一覧を示すと、あまりにも数が多くなるため、ここでは 6 割以上とした。

表 4-6：ソーシャルゲームの作品群において頻繁にみられたゲーム仕様

| カテゴリー | ゲーム仕様の一覧 |
| --- | --- |
| 育成コマンドに関わる仕様 | - |
| 関係値に関わる仕様 | - |
| キャラクターに関わる仕様 | ・ 育成対象キャラは複数体デッキ編成が可能<br>・ 育成対象キャラが複数いる<br>・ 育成対象キャラにはタイプが設定されている<br>・ 育成対象キャラにはレベルがある<br>・ 育成対象キャラには属性が設定されている<br>・ 育成対象キャラのパラメータはメニューから確認できる<br>・ 育成対象キャラの各パラメータは主にバトルで使用される<br>・ 属性は体系化されており同一タイプの育成キャラクターグループがある<br>・ 育成キャラクターには複数のパラメーターがある<br>・ 強化合成に使用した育成対象キャラクターは削除される<br>・ レベルアップすると育成対象キャラのパラメータが上昇 |
| メニュー画面やUIに関わる仕様 | ・ デッキ編成できる育成対象キャラクター数には上限がある<br>・ クエストはリスト表示される<br>・ デッキは複数種類保存できる<br>・ ゲームの最初にプレイヤーキャラクターの名前を決める<br>・ ゲーム内では属性ごとに別々のアイコンで表現されている |
| その他、育成関連仕様 | ・ 育成期間に制限がない<br>・ 強化合成がある<br>・ 次のレベルまでに必要な経験値の値が確認できる |
| その他、戦闘関連仕様 | ・ 属性は任意に変更不可<br>・ プレイヤーキャラクターにはレベルがある<br>ゲームの進行に合わせてクエストが追加される<br>バトルはターン制<br>属性は複数種類存在<br>パラメータとの結びつきが不明な目標がそれとなく提示される |

育成対象キャラに関する仕様はソーシャルゲームにおいて、数多くの仕様が見られる。

一方で、（1）育成コマンドに関わる仕様が必ずしも見られず、キャラの育成は主にカード同士をかけ合わせる強化合成によって行われること、（2）関係値に関わる仕様が必ずしも見られないこと（3）育成期間に制限がない、といった部分が典型的な育成ゲームと比較してゲーム仕様として大きく異なっているということが伺える。

　また、ソーシャルゲームは概ね戦闘ないし試合を表現したものが多数を占めることも特徴として伺える。

### 4.4.4 RPG 群

　同じく、4.3.1（4）で示したロールプレイングゲーム（RPG）の作品群において頻繁にみられたゲーム仕様[21]を表 4-7 に示す。

---

[21] ここでは、6 割以上（3 作品中の 2 作品以上）のものを示した。全体が 3 作品しかないため複数の共通する要素とした場合に 2 作品以上となるためである。

表 4-7：ロールプレイングゲーム（RPG）の作品群において頻繁にみられたゲーム仕様

| カテゴリー | ゲーム仕様の一覧 |
| --- | --- |
| 育成コマンドに関わる仕様 | - |
| 関係値に関わる仕様 | - |
| キャラクターに関わる仕様 | ・タイプは体系化されており同一タイプの育成キャラクターグループがある<br>・育成対象キャラクターが複数いる<br>・プレイヤーキャラクターには育成パラメータがない<br>・育成キャラクターの各パラメータは主にバトルで使用される<br>・育成キャラクターにはタイプがある<br>・育成キャラクターには性別がある<br>・育成キャラクターには複数のパラメーターがある<br>・育成キャラクターには名前とは別にニックネームを任意設定できる<br>・育成キャラクターは戦闘で捕獲して獲得<br>・特性には特定条件で育成キャラクターのパラメーターが上昇するものがある<br>・捕獲できる育成キャラクターの種類はシナリオの進行状況によって追加されていく<br>・スキルポイントの所持数は各育成キャラクターにひもづく<br>・育成キャラクターごとに初期パラメータ及びレベルアップ時の上昇パラメータ値が異なる |
| メニュー画面やUIに関わる仕様 | ・ゲームの最初にプレイヤーキャラクターの名前を決める<br>・自動選択コマンド設定中のキャラはプレイヤーによるコマンド入力がスキップされる<br>・各育成キャラクターのパラメータはメニューから確認可能<br>・パラメータ変動値は装備前・装備後で比較表示される<br>・習得済の効果は白文字で表示 |

| その他育成関連仕様 | ・育成期間に制限がない<br>・どのパラメータが上昇するかはプレイヤーは任意で選択できない<br>・レベルアップするとパラメータが上昇する<br>・セーブデータは1つのみ作成できる<br>・デッキ編成数には上限がある<br>・レベルアップ時に獲得できるスキルポイントの値はレベル帯によって異なる<br>・スキルポイントの振り分けはレベルアップ時及びメニューから振り分け画面を表示できる<br>・各スキルはスキルポイントを振り分けて強化<br>・振り分け画面では各スキルを仮選択して左右ボタン入力で1ポイントずつ付与数を増減できる<br>・振り分け画面には各スキルのポイント付与累積数が表示<br>・振り分け画面には振り分け可能なスキルリスト所持スキルポイント合計数現在のパラメータが表示<br>・特定どうぐを使用するとどうぐ名に対応したパラメータが上昇<br>・付与数に応じて所持スキルポイント合計数が変動する<br>・各スキルのスキルアップで獲得できる効果及び必要スキルポイントがリスト表示されている<br>・選択種類に沿ってAIがコマンドを自動選択するコマンドがある |
|---|---|
| その他、戦闘関連仕様 | ・バトルでは攻撃以外にもせんとうを離脱するコマンドがある<br>・タイプは複数種類存在<br>・タイプは任意に変更不可<br>・ターン開始時に味方のコマンドを1キャラずつ選択する<br>・味方キャラのコマンド入力がすべて完了すると敵味方が行動を開始<br>・「こうげき力」パラメータは物理属性の攻撃でダメージ量に影響<br>・「すばやさ」パラメータはバトル中の行動順に影響<br>・バトルは敵味方どちらか全員のHPが0になると終了<br>・マップ移動およびイベント中にバトルが発生する<br>・リスポーンの際は所持ゴールド減少ペナルティが発生<br>・未習得の効果はグレーアウトで表示<br>・敵全員のHPが0になった場合は勝利扱いとなり報酬を獲得しプレイが再開する<br>・特性は複数種類存在<br>・自動選択コマンドはバトル中以外でもメニュー画面から設定可能<br>・装備後の値が装備前より高い場合と低い場合でテキストカラーが変化する<br>・「しゅび力」パラメータは被ダメージ量に影響<br>・バトルはターン制で進行<br>・戦闘で捕獲する場合対象のHPが低いほど成功率が上昇する<br>・全員のHPを最大まで回復するコマンド実行中にMPが足りない場合は足りなくなった時点で終了<br>・装備数には上限がある<br>・装備変更はメニューからいつでも可能<br>・装備するとパラメータが変動する |

　育成対象キャラに関する仕様については、ソーシャルゲーム群と同様に、RPG群においても、数多くの共通する仕様が見られる。

大きく異なっている点としては（１）育成コマンドに関わる仕様が必ずしも見られず、育成対象キャラの成長は主に敵キャラクターとのバトルを通して得られる経験値の蓄積（およびそれに伴うレベルアップ）によってなされる（２）関係値に関わる要素が必ずしもないこと（３）育成期間に制限がないといった特徴がみられる。

　一方で、RPG群の作品は、ゲーム全体がさまざまな要素が多数結びついたものであるため、そもそものゲーム仕様の数自体が多く、とりわけ戦闘に関わる仕様の数が多いといった特徴が伺える。

### 4.4.5 境界例群

　同じく、4.3.1（5）で示した育成ゲームの境界例群において頻繁にみられたゲーム仕様[22]を表4-8に示す。

---

[22] ここでは、10割以上（2作品中の2作品以上）のものを示した。全体が2作品しかないため複数の共通する要素を抽出する場合に必然的に2作品以上となるためである。

表 4-8：境界例群において頻繁にみられたゲーム仕様

| カテゴリー | ゲーム仕様の一覧 |
|---|---|
| 育成コマンドに関わる仕様 | ・練習コマンドは育成期間内であれば無制限に実行可能<br>・育成コマンドには成功失敗がある<br>・パラメータ上限に達している場合は育成コマンドを実行しても無効となる |
| 関係値に関わる仕様 | - |
| キャラクターに関わる仕様 | ・育成対象キャラクターが複数いる |
| メニュー画面やUIに関わる仕様 | ・システムメニューがある<br>・セーブデータは1つのみ作成できる<br>・メイン画面には現在の時間が表示されている<br>・メイン画面には表示中の育成キャラクター名が表示されている<br>・パラメータ確認画面に表示されないパラメータがある<br>・ゲームはゲーム内における24時間が自動経過する |
| その他育成関連仕様 | ・育成パラメーターを競うコンテンツがある<br>・種目ごとに使用するパラメータが異なる |

境界例群においては、関係値以外の仕様では、典型的育成ゲーム群と共通する仕様がいくつか見られ、典型的な育成ゲームの仕様の影響がある程度存在しているであろうことが伺える。特に複数の育成コマンドを選択するという仕様は、典型的な育成ゲームとの間に共通性が見られる。

一方で、大きな違いとしては、（1）必ずしも関係値がないことや、（2）必ずしも育成対象キャラクターが複数いるわけではないといった点が挙げられる。

## 4.5 個別のゲーム仕様について

ここまでの全体傾向において、育成コマンドに関連する仕様の有無、登場キャラクター間の関係値に関わるゲーム仕様、育成対象キャラクターに関わる仕様、その他の育成関連仕様、バトル仕様等のそれぞれの仕様において5つそれぞれの群において差異があることが示された。

本節では、個別のゲーム仕様がどのように本調査の対象となった作品に見られているのか、その傾向について、具体的にゲーム画面や攻略本等の資料を引用しながら記述していく。な

お、この項での説明は、それぞれの個別の仕様の例示である。網羅的に説明し、当該仕様の最初の起源を示すものではない。

この節では前節までのデータに基づき、下記のゲーム仕様の具体的な具体的な実現方法について示していく。

- ● 複数ある育成コマンドを選択しながらゲームが進行する
- ● 育成コマンド内にさらに複数種類のコマンドがある
- ● コスト系パラメーターを回復するコマンドがある
- ● 登場キャラを任意で設定可能(ex:育成開始時に編成/育成中に仲間にする、など)
- ● イベントが発生すると、育成キャラのパラメーターが増減
- ● 各登場キャラに応じた固有イベントが発生
- ● 育成キャラには基礎パラメーター以外の能力（スキル）がある
- ● 育成パートと、育成したキャラを活用できるパートがある
- ● 育成キャラと登場キャラの間の関係値に応じたイベント発生
- ● 抽選で育成対象キャラクター等のゲーム内コンテンツが入手できる
- ● 育成対象キャラ等のゲーム内コンテンツにレアリティがある

## 4.5.1 複数ある育成コマンドを選択しながらゲームが進行する

『プリンセスメーカー』等の育成ゲームにおいては一般に、体力、腕力、知力等のパラメーターが育成対象キャラに設定されている。育成ゲームにおいては、どのような育成対象キャラをどのような方向性で育成していくのかを考えるということがゲームの体験の中心的な部分を占めており、どのパラメーターを特に育成したいかに応じて、複数ある育成コマンドのうちからターン毎に育成コマンドを選択するという仕様が、多くの育成ゲームに見られる。

この仕様は「育成ゲーム」ジャンル全体の中でも特に根幹的な仕様の一つとなっており、典型的な育成ゲーム群においては、ほぼこの仕様が実装されている。

一方で、育成ゲームジャンルと異なると思われるタイトルとして調査した『ドラゴンクエストIX』のようなRPG群や、『FGO』のようなソーシャルゲーム群のタイトルにおいては、こ

のゲーム仕様の存在は一般的ではなく、育成ゲームジャンルに特有かつ根幹的な要素をなすゲーム仕様であることが伺える。

『プリンセスメーカー』　　　　　　　　　　　　　　　　　　ダービースタリオン

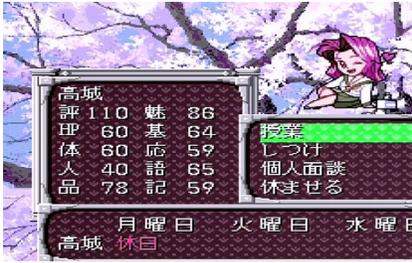
『卒業』

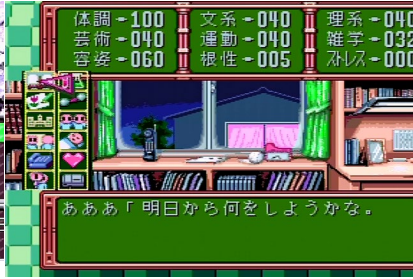
『ときめきメモリアル』

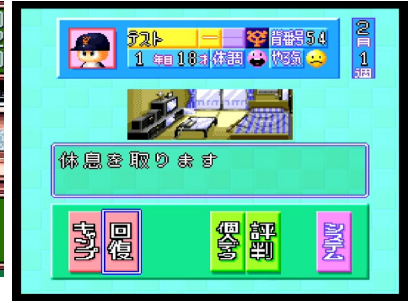
『パワプロ3』サクセスモード

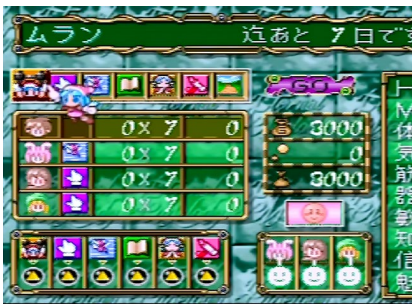
『エターナルメロディ』

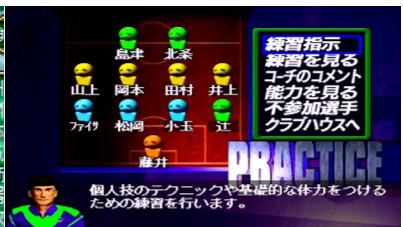
『サカつく』

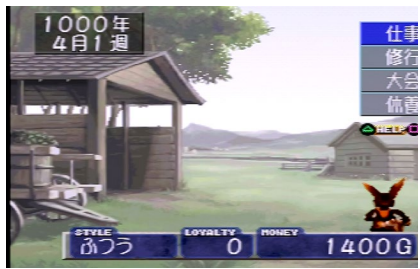
『モンスターファーム』

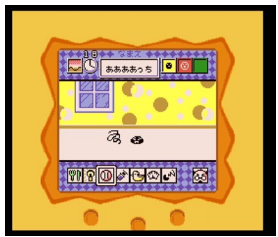
『たまごっち』

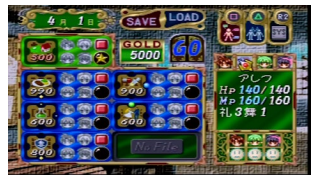
『悠久幻想曲』

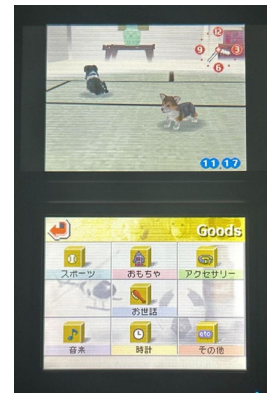
『nintendogs』

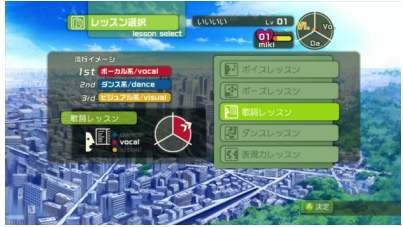
『アイドルマスター』

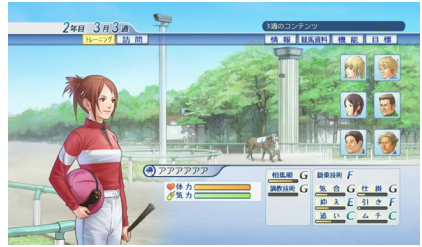
『ウイニングポスト

ワールド』

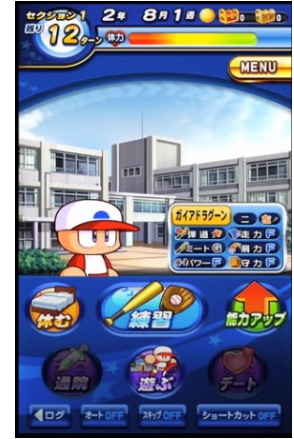
『パワプロアプリ』

図 4-8：各種ゲームにおける育成コマンド

### 4.5.2 育成コマンド内にさらに複数種類のコマンドがある

　育成コマンドの中にさらに階層的に複数の育成コマンドが存在しているという仕様も、育成ゲームジャンルにおけるかなり中心的な仕様だと言える。初期作品である『プリンセスメーカー』と『ダービースタリオン』の双方において、すでにこの仕様が確認できる。

図 4-9：『プリンセスメーカー』における育成コマンドの階層

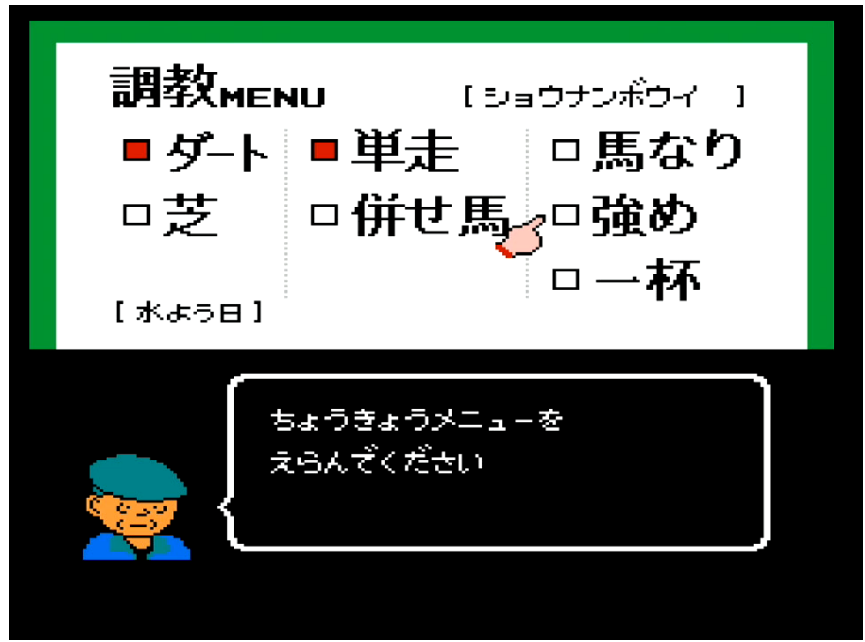

図 4-10：『ダービースタリオン』における育成コマンドの階層

この要素についても、後続する典型的な育成ゲーム群の多くの作品において確認することができる。作品によっては、育成ゲームにおける必須の仕様とまでは言えないものの、かなり一般的に見られる仕様であると言ってよいだろう。

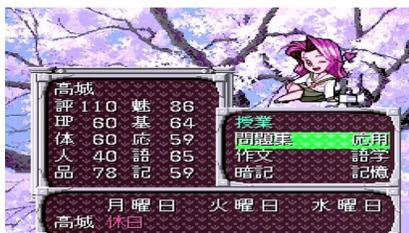
『卒業』

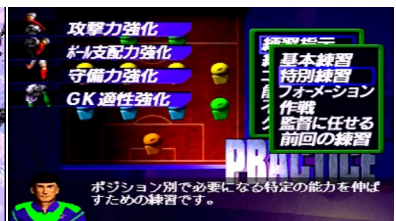
『サカつく』

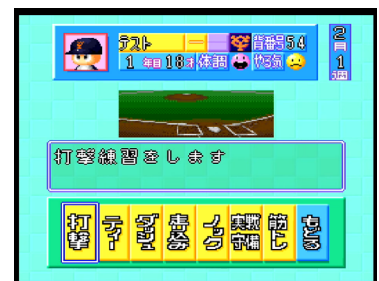
『パワプロ3』サクセスモード

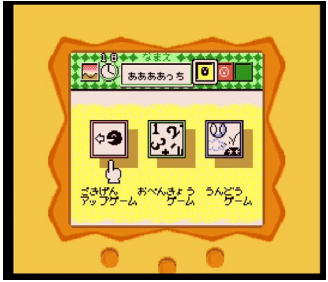
『たまごっち』

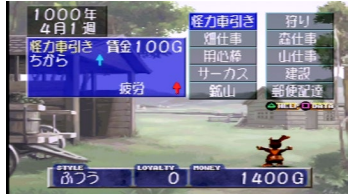
『モンスターファーム』

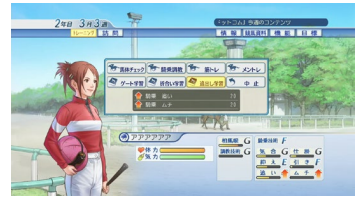
『ウイニングポスト ワールド』

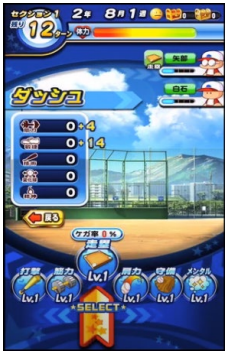
『パワプロアプリ』

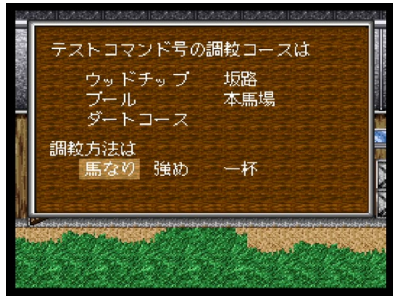
『ウイニングポスト 2』

図 4-11：各種ゲームにおける育成コマンドの階層化

### 4.5.3 コスト系パラメーターを回復するコマンドがある

休憩、休息などのコスト回復コマンドといったゲーム仕様も、典型的育成ゲーム群において、比較的初期から確認できるゲーム仕様の一つである。

ここで言うコストとは、他の育成コマンドを実行するために必要な資源として機能するものであり、コマンドを実行すると消費するものを指す。ただし、これは多く場合、金銭のような交換可能な資源としてではなく、キャラクターの状態に紐づいた体力（HP）などの値が用いられることが多い。図 4-12 に『卒業』における例を示す。

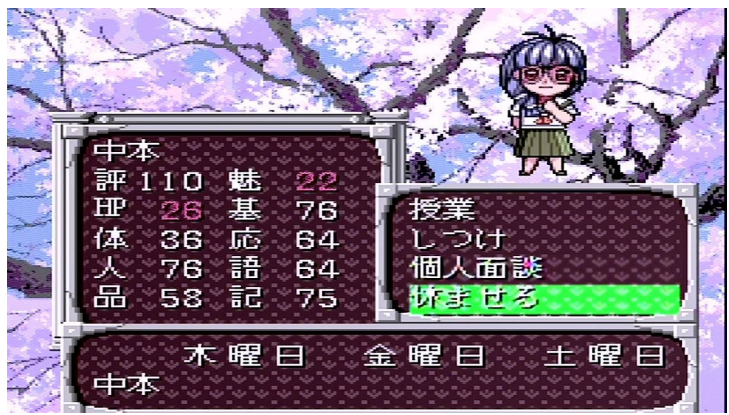

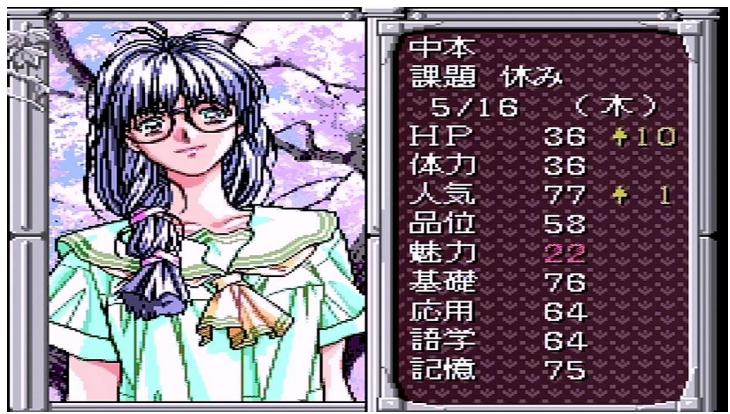

図 4-12：『卒業』にて「休ませる」コマンドを選択

コスト系パラメータ（ここでは HP）が回復している。

育成ゲームの初期のタイトルである PC-98 版『プリンセスメーカー』、ファミリーコンピュータ版『ダービースタリオン』からその存在を確認することができる。

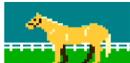

図 4-13：『プリンセスメーカー』コスト系パラメーター（疲労）

図 4-14：『ダービースタリオン』のコスト系パラメーター

育成している競走馬をやすませるコマンドを実行したところ [23]

---

[23] ただし、『ダービースタリオン』においては、パラメーターの回復があったことは直接に目視で確認はできない。

このゲーム仕様もほとんどの後続する育成ゲームにおいて一般的に観察される仕様であり、育成ゲームジャンル自体の初期からの根幹に近い要素の一つと見なすことができるだろう。

ただし、ここで何が「コスト系パラメーター」として機能しているのかについては作品に応じて、多様な名称がある。典型的に用いられる「体力」や「HP」のほかに「ストレス（『ときめきメモリアル』）、気力（『ウイニングポスト』）、お腹の減り具合（『たまごっち』『nintendogs』）、喉の乾き具合（『nintendogs』）などがある。

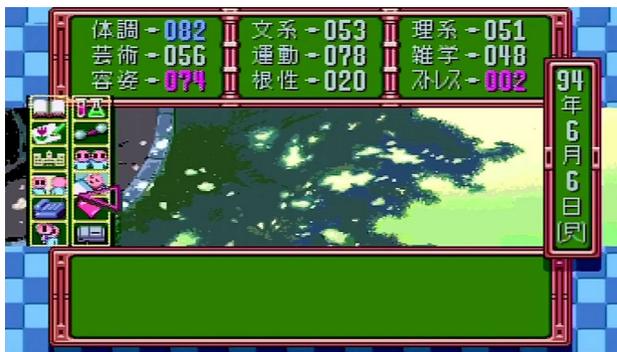
『ときめきメモリアル』

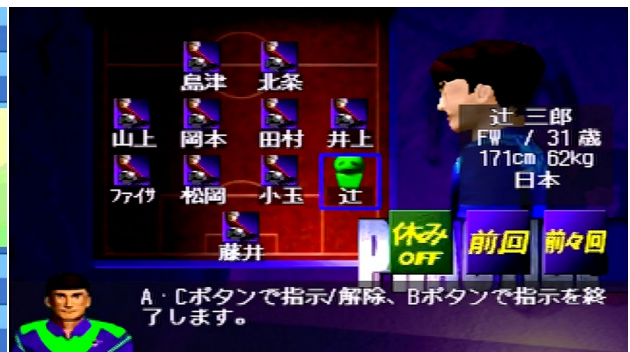
『サカつく』

（ここでは「ストレス」がコマンドによって回復している）

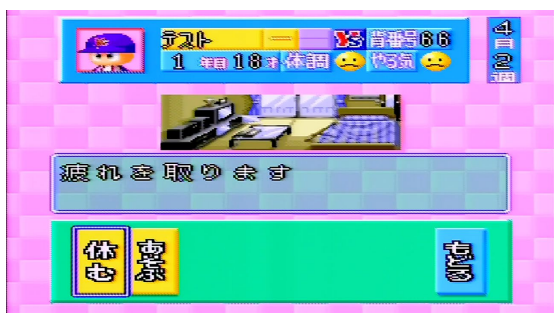
『パワプロ3』サクセスモード

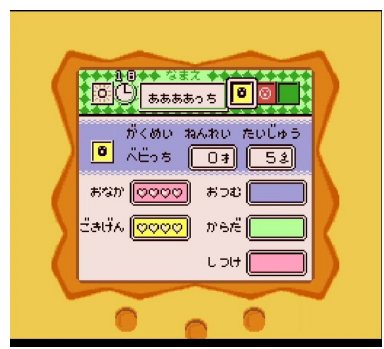

『たまごっち』（「おなか」をどのぐらい空かせているかと、「ごきげん」の良さがコスト系パラメーターとして機能している）

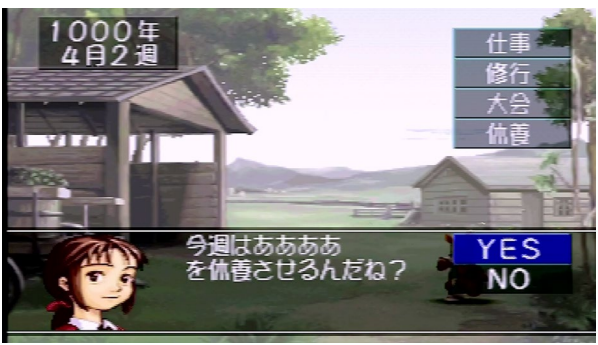

『モンスターファーム』

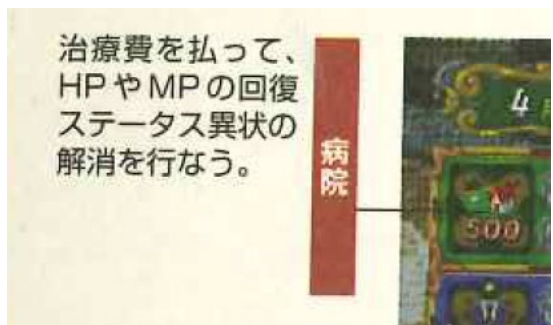

『悠久幻想曲』攻略本 p.10 より

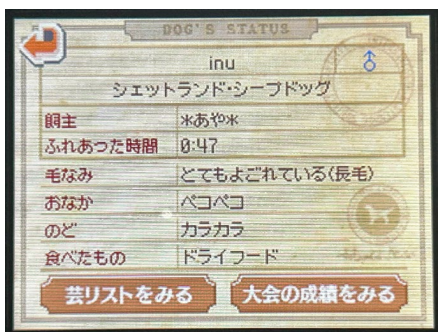

『nintendogs』（「のど」の乾き具合と、「おなか」の空き具合がコスト系パラメーターとして機能している）

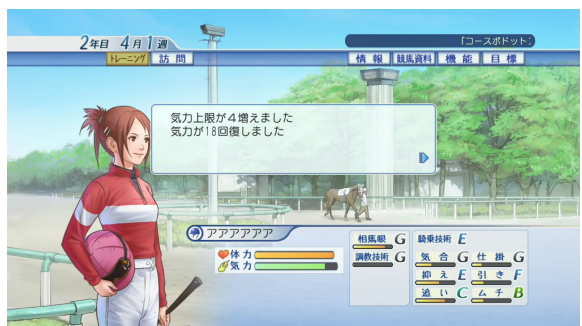

『ウイニングポストワールド』（気力）

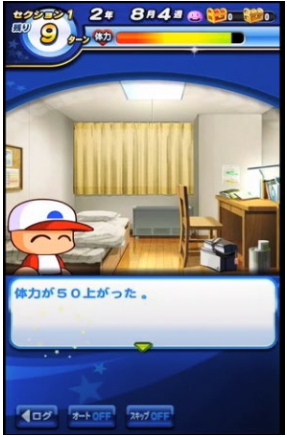
『パワプロアプリ』

図 4-15：各ゲームのコスト系パラメーター

　もっとも、育成ゲーム以外のジャンルにおいても「体力」や「HP」のようなパラメーターは広く用いられており、『ドラゴンクエスト』（1986, エニックス, ファミリーコンピュータ）のような RPG においても「宿屋」に泊まることで HP を回復させることができるが、育成ゲームジャンルにおけるコスト系パラメーターに特徴的なのは、当該のパラメーターが育成のためのコマンドを実行することで消費されるものとして設定されているという点である。この点において、育成ゲームのコスト系パラメーターの存在は、RPG などにおける一般的なゲーム仕様とは区別可能なことが多い。育成ゲームジャンルにおける根幹的な特徴をなす仕様の一つであると言えよう。

**4.5.4 登場キャラを任意で設定可能(ex:育成開始時に編成/育成中に仲間にする、など)**

　ゲーム内において、ゲームプレイヤーが主人公以外の登場キャラクターを任意に登場させることを選ぶことのできる仕様は、育成ゲームジャンルに限らず、幅広いゲームにおいて観察されるゲーム仕様である。

下記、『ポケットモンスター ルビー・サファイア』の例を示す。この作品では、シリーズの最初期（『ポケットモンスター 赤・緑』）から見られるゲーム仕様として、図 4-16 のように最初に育成対象とするゲーム内キャラをプレイヤーが三択の中から選ぶことができる。

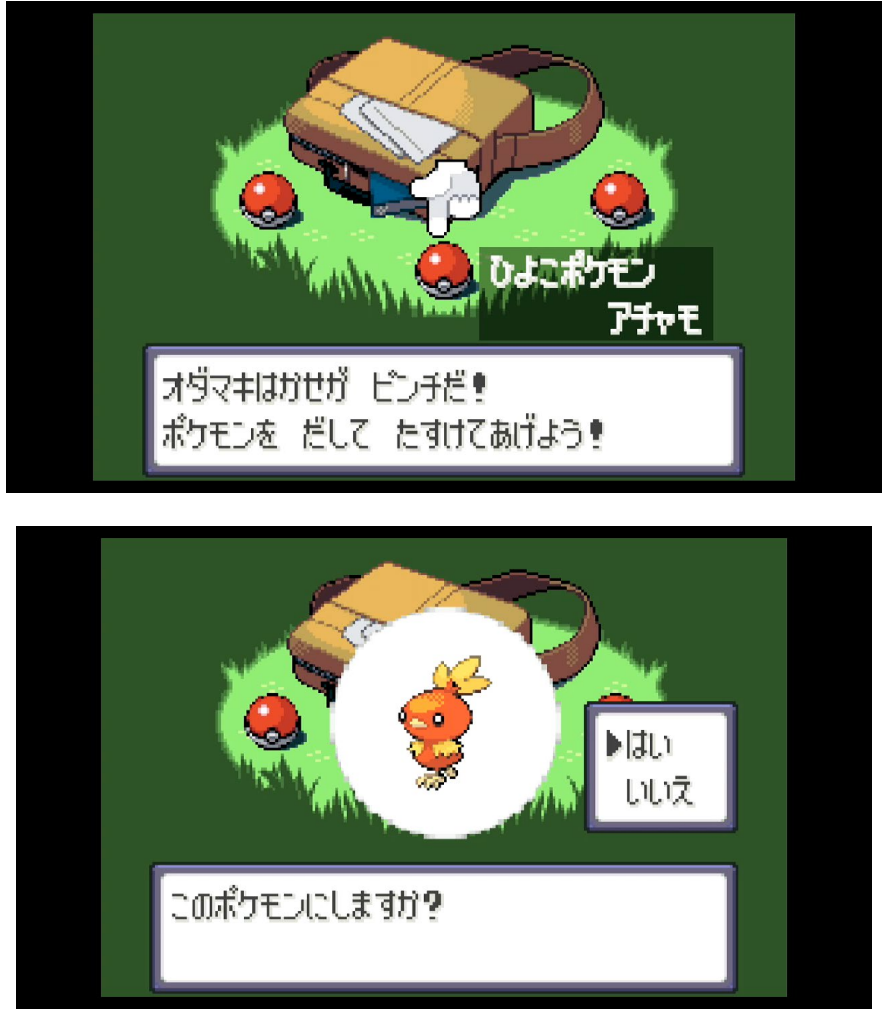

図 4-16：『ポケモン　ルビー・サファイア』の登場キャラの選択シーン

　このゲーム仕様は、『ポケットモンスター　ルビー・サファイア』の攻略本の中にも明記されており、『ポケットモンスター』シリーズを遊んだことのある、多くのプレイヤーにとって、よく知られた仕様であると言える。

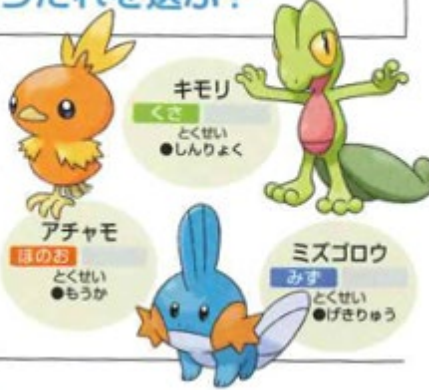

図 4-17：『ポケモン　ルビー・サファイア』の登場キャラの選択

元宮＆ワンナップ編著（2002, p.35）より引用

　育成ゲームにおけるこうした仕様の例としては、例えば『ウイニングポストワールド』における育成対象以外の登場キャラクターの選択をする場面などに、共通するゲーム仕様を見出すことができる。

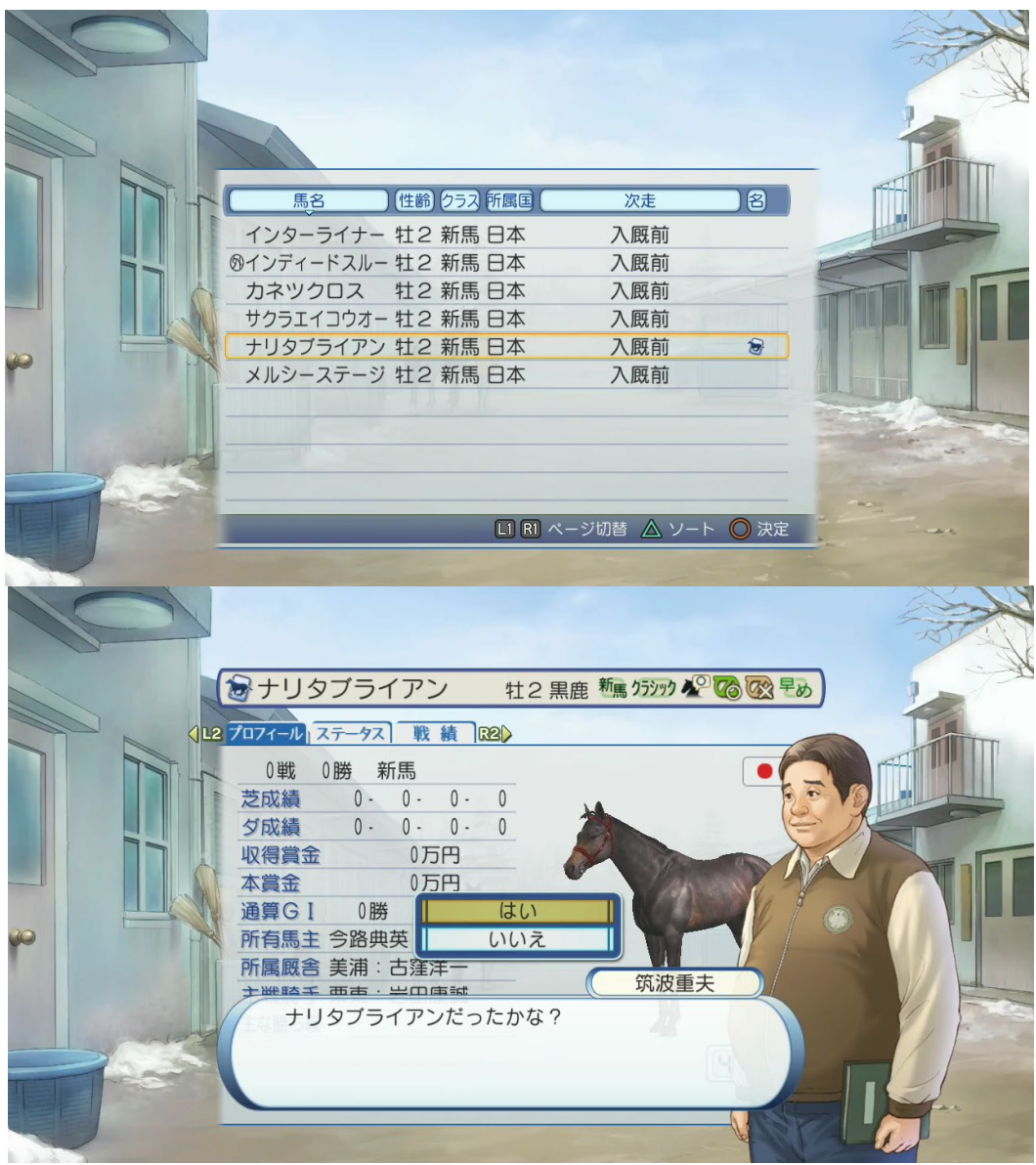

図 4-18:『ウイニングポストワールド』における育成対象以外の登場キャラ選択場面

図 4-19：『ウイニングポスト 2』の登場キャラの選択

『ウイニングポスト 2』パッケージに付属の説明書（pp.12-13）より引用

さらに、『サカつく』におけるスポンサーの契約や、『エターナルメロディ』『悠久幻想曲』における仕様に見出すことができる。

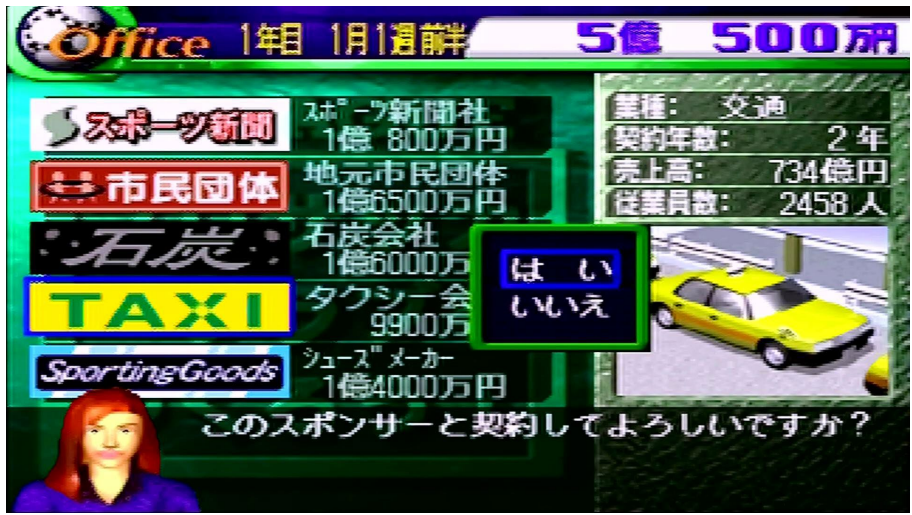

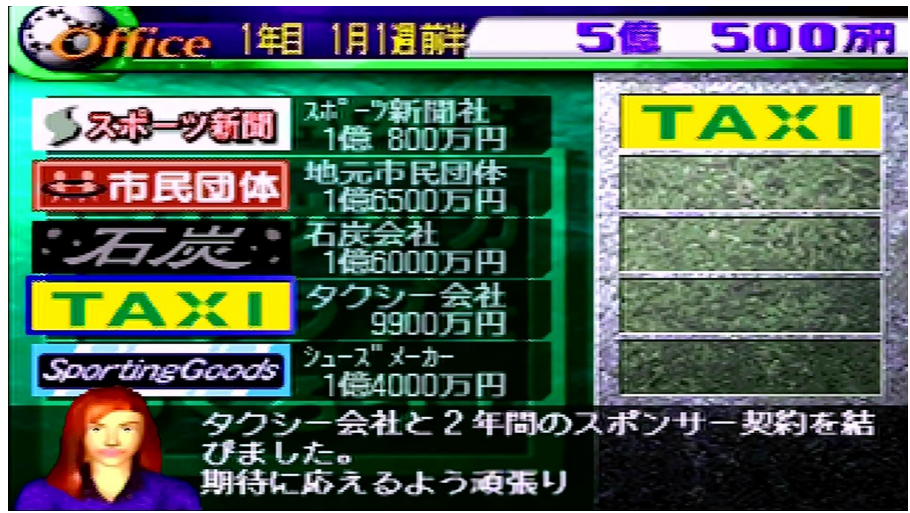

図 4-20：『サカつく』におけるスポンサーの契約

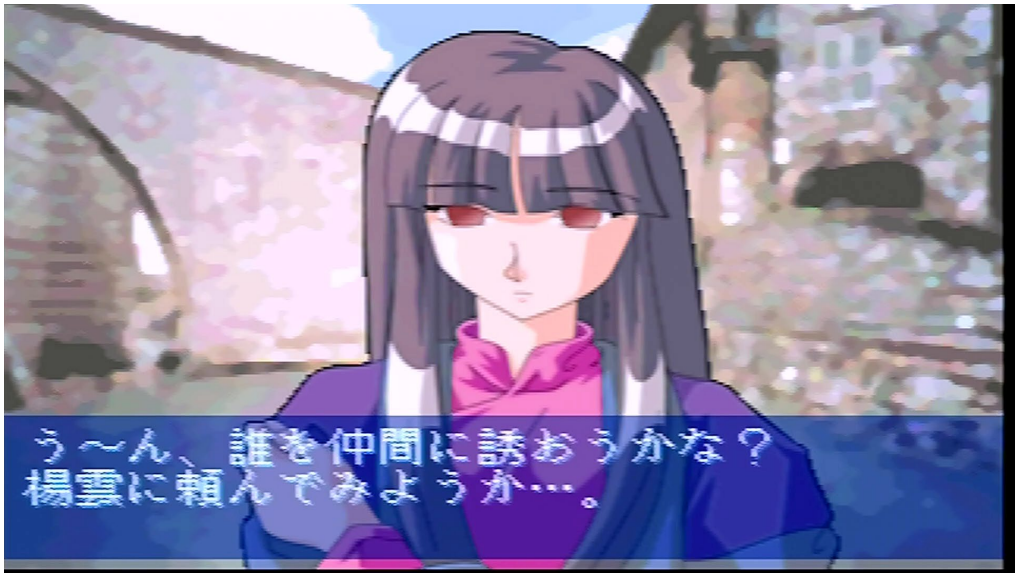

図 4-21：『エターナルメロディ』における任意キャラの選択

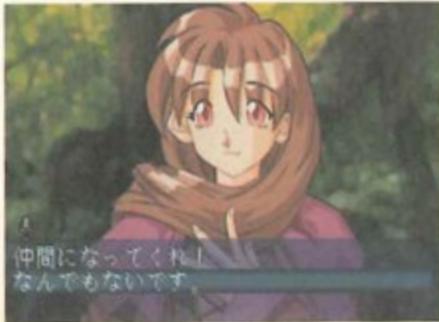

図 4-22：『エターナルメロディ』における任意キャラの選択

電撃 PlayStation＆電撃 SEGAEX 編集部（1996, p.7）より引用

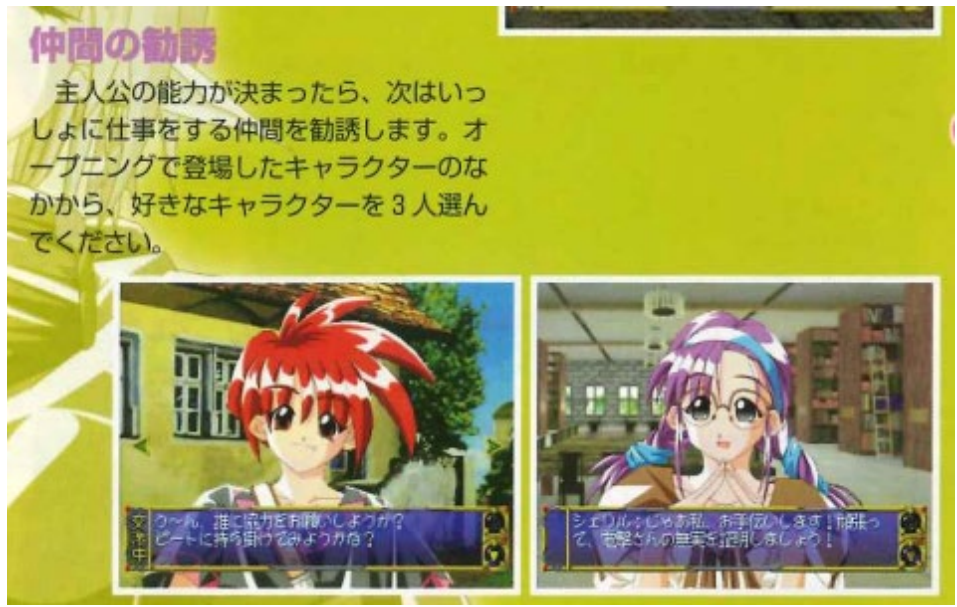

図 4-23：『悠久幻想曲』における任意キャラの選択

『悠久幻想曲』パッケージに付属の説明書（p.5）より引用

　先述したとおり、この仕様はいわゆる「育成ゲーム」というジャンル名で名指される作品にとどまらず、ゲーム内のキャラクターを育成する要素をもった幅広いゲームにおいて見出すことのできるゲーム仕様である。下記『ドラゴンクエストモンスターズジョーカー 2』や、『パズル&ドラゴンズ』などにおいてもこの仕様が実装されていることを確認できる。

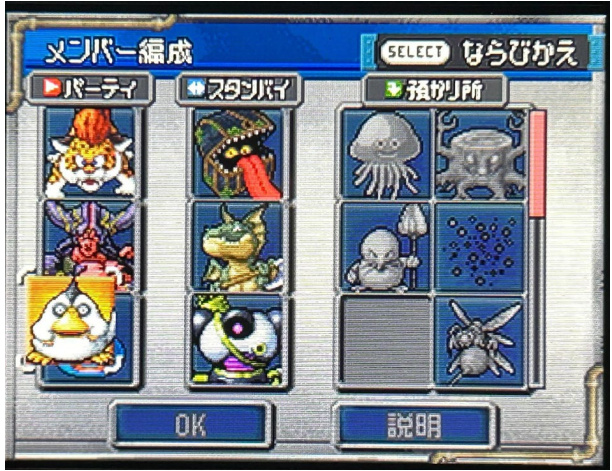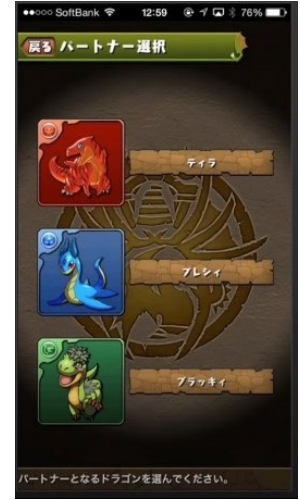

『ドラゴンクエストモンスターズジョーカー 2』　　　『パズドラ』

図 4-24：任意キャラの選択

### 4.5.5 イベントが発生すると、育成キャラのパラメーターが増減

育成ゲームには、ゲーム内の物語設定の文脈等に呼応して、キャラクターの成長が促されるようなイベントが設定されていることがしばしば見られる

たとえば、下記に示す『エターナルメロディ』攻略本 p.68 では、多数のゲームプレイ上で多数のイベントが発生し、それぞれのイベントの中で適切な選択肢を選びとることで、育成対象キャラクターのパラメーターが成長するゲーム仕様となっている。

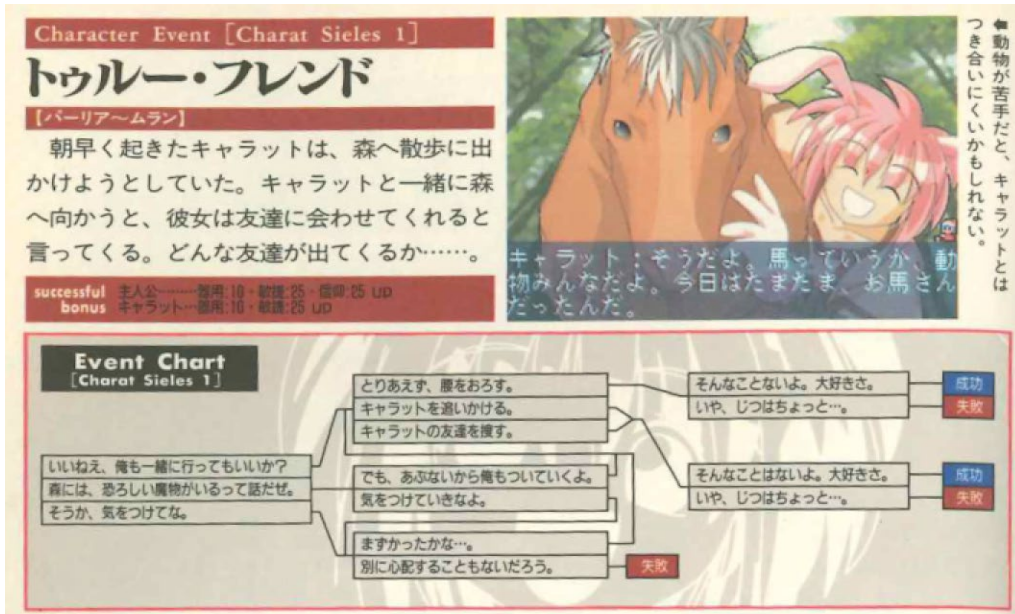

図 4-25：『エターナルメロディ』におけるパラメーター成長の成功可否の解説

電撃 PlayStation＆電撃 SEGAEX 編集部（1996, p.68）より引用

このイベントによってパラメーターが増減される仕様の実現方法には複数の種類のものがある。

（1）イベントの発生条件：イベントの発生の仕方が確率的に発生するものと、一定の条件を満たすと発生するものかで違いがある。

（2）パラメーター上下の条件：また、イベントが発生したら自動的にパラメーターが上昇するものと、イベントが発生したのちプレイヤーの対処の良し悪しに（適切なコマンドの入力ができたかどうかなど）よってパラメーターの上下の度合いが変わるものがある。

どのような実装方法が用いられるかについては、ある程度のバリエーションが見られるが、何かしらの条件によってイベントが発生し、それに応じて育成キャラのパラメーターが増減するという仕様は、育成ゲームにおいて、一般に観察することのできるゲーム仕様の一つである。

こうしたゲーム仕様は、育成ゲームにおいて 90 年代中盤には一般に見られる仕様となっている。下記、『ときめきメモリアル』、『子育てクイズマイエンジェル』などに確認できる。

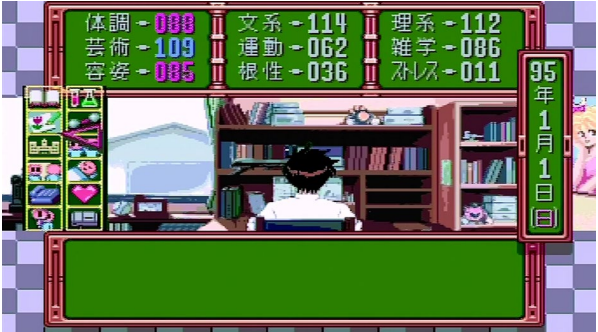

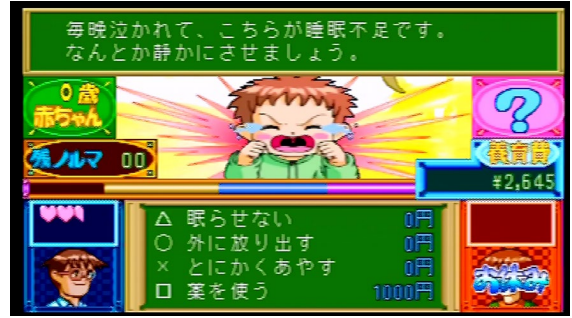

『子育てクイズマイエンジェル』

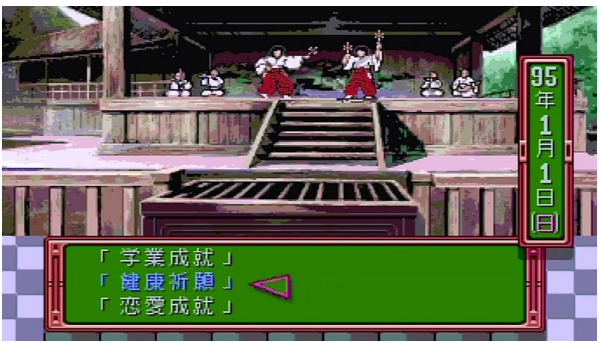

『ときめきメモリアル』

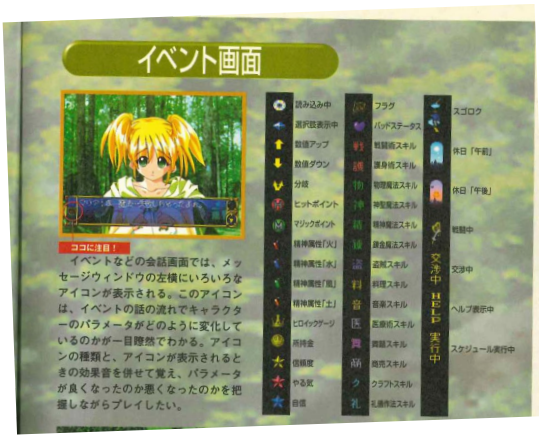

『悠久幻想曲』 攻略本 p.13

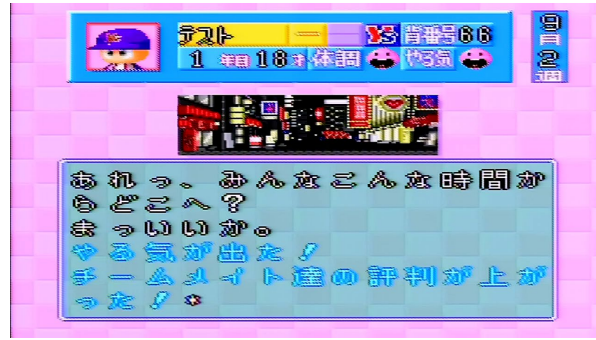

『パワプロ 3』 サクセスモード

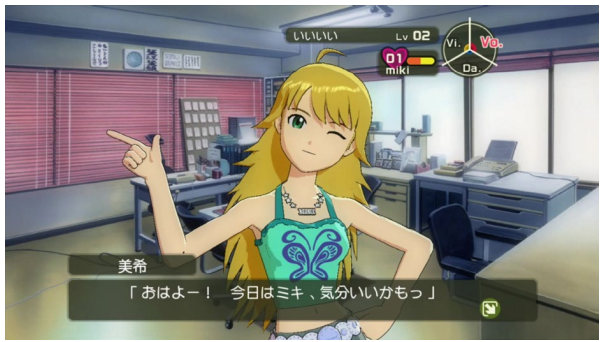

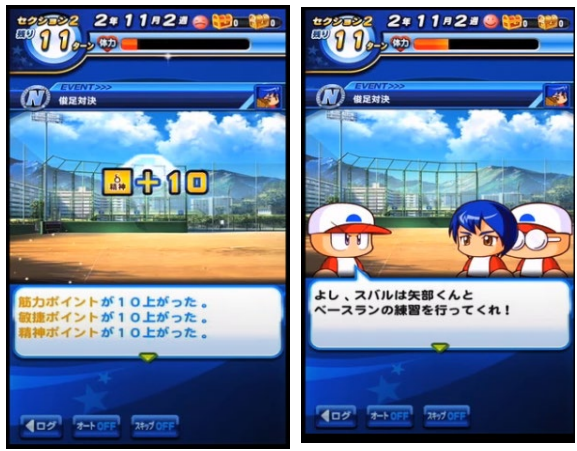

『パワプロアプリ』

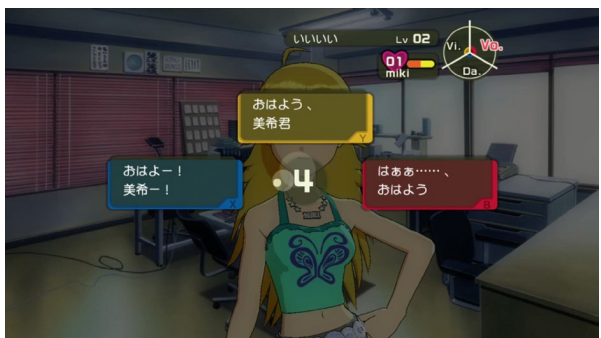

『アイドルマスター』

図 4-26：各種ゲームにおけるパラメーター増減

### 4.5.6 各登場キャラに応じた固有イベントが発生

キャラに固有のイベントが設定されている、というゲーム仕様は、様々な設定をもったキャラクターが多数登場することをゲームの主たる魅力とする作品において、一般的に見られるゲーム仕様である。

たとえば、1992 年に発売された恋愛アドベンチャーゲーム『同級生』などは、各登場キャラクターに豊富な固有イベントが用意されていることで知られる作品も存在している。育成ゲームに限らず、ゲーム内のキャラクターの存在が物語上あるいはゲームプレイ上において重要な魅力の要素となっている作品において存在するゲーム仕様である。

育成ゲームの中では、多数のキャラクターが育成対象として登場するゲームにおいては、しばしば見られる仕様である。先述の『エターナルメロディ』においても、キャラ別のイベントが割り振られており、下記に示すとおり攻略本の中で「キャラット・シールズ」、「ティナ・ハーヴェル」など、キャラクターごとの固有イベントが紹介されているページがあることからも、その存在を確認できる。

![イベント紹介 キャラット・シールズ Character Event [Charat Sieles 1] トゥルー・フレンド【パーリア〜ムラン】 朝早く起きたキャラットは、森へ散歩に出かけようとしていた。キャラットと一緒に森へ向かうと、彼女は友達に会わせてくれると言ってくる。どんな友達が出てくるか……。successful bonus 主人公……器用:10・敏捷:25・信仰:25 UP キャラット……器用:10・敏捷:25 UP　キャラット：そうだよ、馬っていうか、動物みんなだよ。今日はたまたま、お馬さんだったんだ。　動物が苦手だと、キャラットとはつき合いにくいかもしれない。]

図 4-27：『エターナルメロディ』におけるイベントの一例
電撃 PlayStation＆電撃 SEGAEX 編集部（1996, p.68）より引用（上）
同前書(p.74) より引用（下）

図 4-28：『悠久幻想曲』におけるイベントの一例

電撃 PlayStation 編集部＆電撃 SEGASATURN 編集部（1997, p.66）より引用

　育成ゲームというジャンルのなかでもすべての作品において確認できるというゲーム仕様ではないが、特にキャラクターを数多く登場させるような育成ゲームのゲーム仕様としては頻繁に見られる仕様であると言える。

図 4-29：『ウイニングポスト 2』における特定キャラのイベント発生条件

キャラクターの裕木江奈の固有イベントについての解説がある。

『ウイニングポスト 2』攻略本（サラブレッド探偵局 1995, p.101）より引用

### 4.5.7 育成キャラには基礎パラメーター以外の能力（スキル）がある

　育成ゲーム作品のなかでも、戦闘や試合といった育成パートの結果を活用するパートを備えた作品の中では、ごく一般的な仕様である。

ここでいう「スキル」とは、ゲーム内の設定としてはキャラクターのもつ特殊能力（魔法、戦闘技、特技、超能力など）として説明されるものである。ゲームの技術的な仕様としては、表示されているパラメータ（攻撃力など）を介して通常時に行われる出力よりも、追加的な効果を付与されるもの全般を指す。

スキルのなかには様々なものがあり、（1）スキルを覚えたら常時その効果が追加されるもの[24]、（2）コマンドを選ぶことで発動するもの[25]、（3）何らかの資源を消費することで発動するもの[26]など細かく区分すれば様々なスキルがあるが、ここではそれらを総称してスキルと呼ぶ。

育成ゲームに限らず、『ドラゴンクエストIX』のような戦闘パートのようなゲーム仕様を備える作品においては、一般に見られる仕様であるといえる。

下記、『エターナルメロディ』の例を示す。

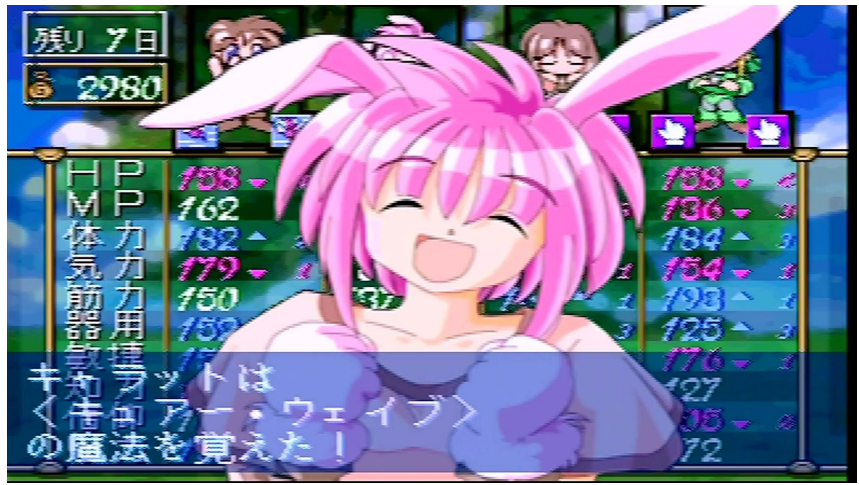

---

[24] 基礎攻撃力が1割アップするものや、ターン毎のHPやMPの回復など。パッシブスキルや、常時発動スキルと呼ばれることが多い。
[25] 複数の敵に一度だけ攻撃するものや、コマンドを選んだターンでHPの回復が行われるものなど。アクティブスキルと呼ばれることが多い。
[26] MPやターン数、気力、HPなどを消費するものが多い。典型的なのはキャラクターのMPを消費して魔法を発動するもの。アクティブスキルの一種であることが多いが、パッシブスキルにおいて資源を消費し続けるタイプのものもある。

図 4-30：『エターナルメロディ』における「魔法」スキルの取得

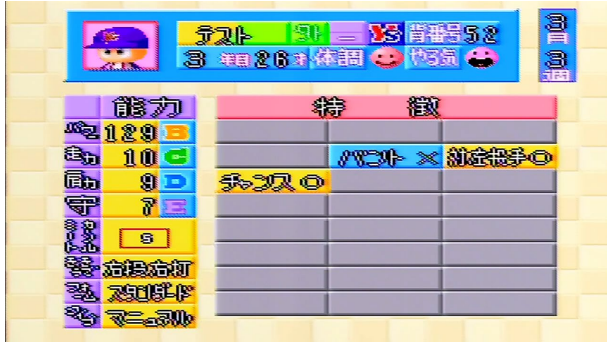 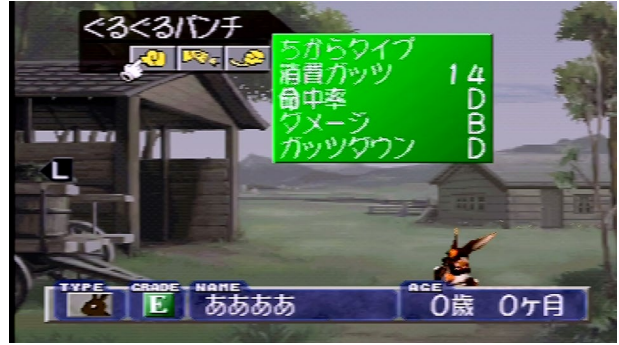

『パワプロ 3』サクセスモード　　　　　　　『モンスターファーム』

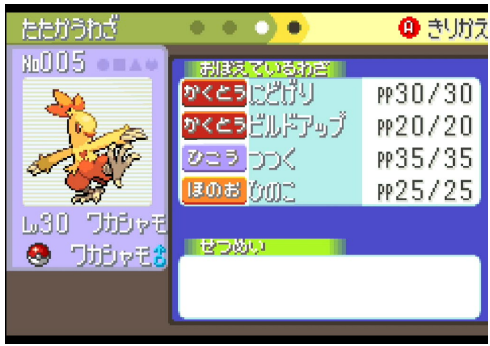 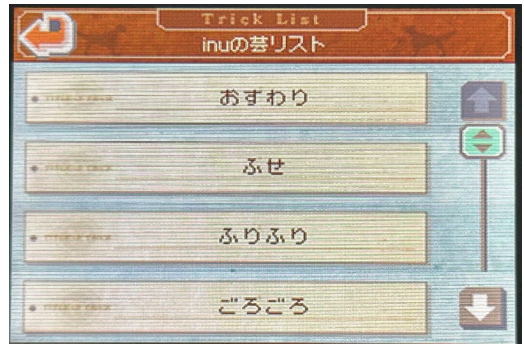

『ポケモン　ルビー・サファイア』　　　　　『nintendogs』

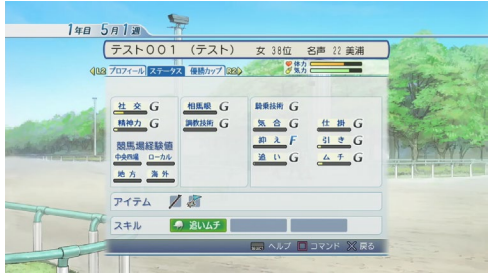

『ウイニングポストワールド』

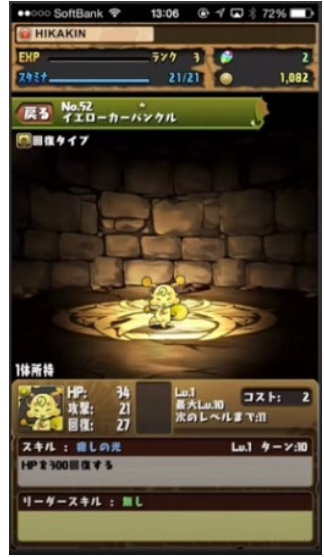

『パズドラ』

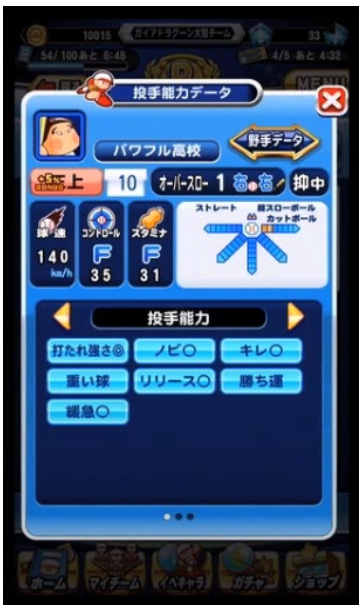

『パワプロアプリ』

図 4-31：その他ゲームにおけるスキルの取得

## 4.5.8 育成パートと、育成したキャラを活用できるパートがある

育成ゲームというジャンルの性質上、育成した内容の結果が、何らかの形で成功、あるいは失敗を評価されるゲーム仕様が存在することが、このジャンルをゲームたらしめる根幹をなしていると考えられる。

この成功と失敗の評価をゲームプレイヤーにフィードバックするゲーム仕様として、主たる実装方法としては（１）ゲームのどこかで成功か失敗の結果だけが知らされるというゲーム仕様である場合と、（２）「通常の育成コマンドを通じたパラメーターの上下を行う育成パート」と「育成したキャラの育成した結果を活用するパート」の二つのゲームパートを同時に存在させることで、育成の成果を体感させるといったものになる。

双方の仕様は、一つの作品の中で、どちらかだけが存在するという形式のものではなく、双方を実装している作品が、初期から多数見受けられる。

『プリンセスメーカー』では、第一に、ゲーム全体が終わった後に育成対象キャラクターである娘がどのような職業につくのか、によって育成結果に対するフィードバックが返ってくるゲーム仕様がある。

第二に、同時に、育成したキャラクターのパラメーターによって、育成キャラクターを戦わせて操作することのできる仕様が存在する。これは一般的な RPG の戦闘と同様のパートとして実装され、「武者修行」のパートとして通常のコマンドによる育成パートとは別に存在している。

下記に、武者修行パートの図を示す。

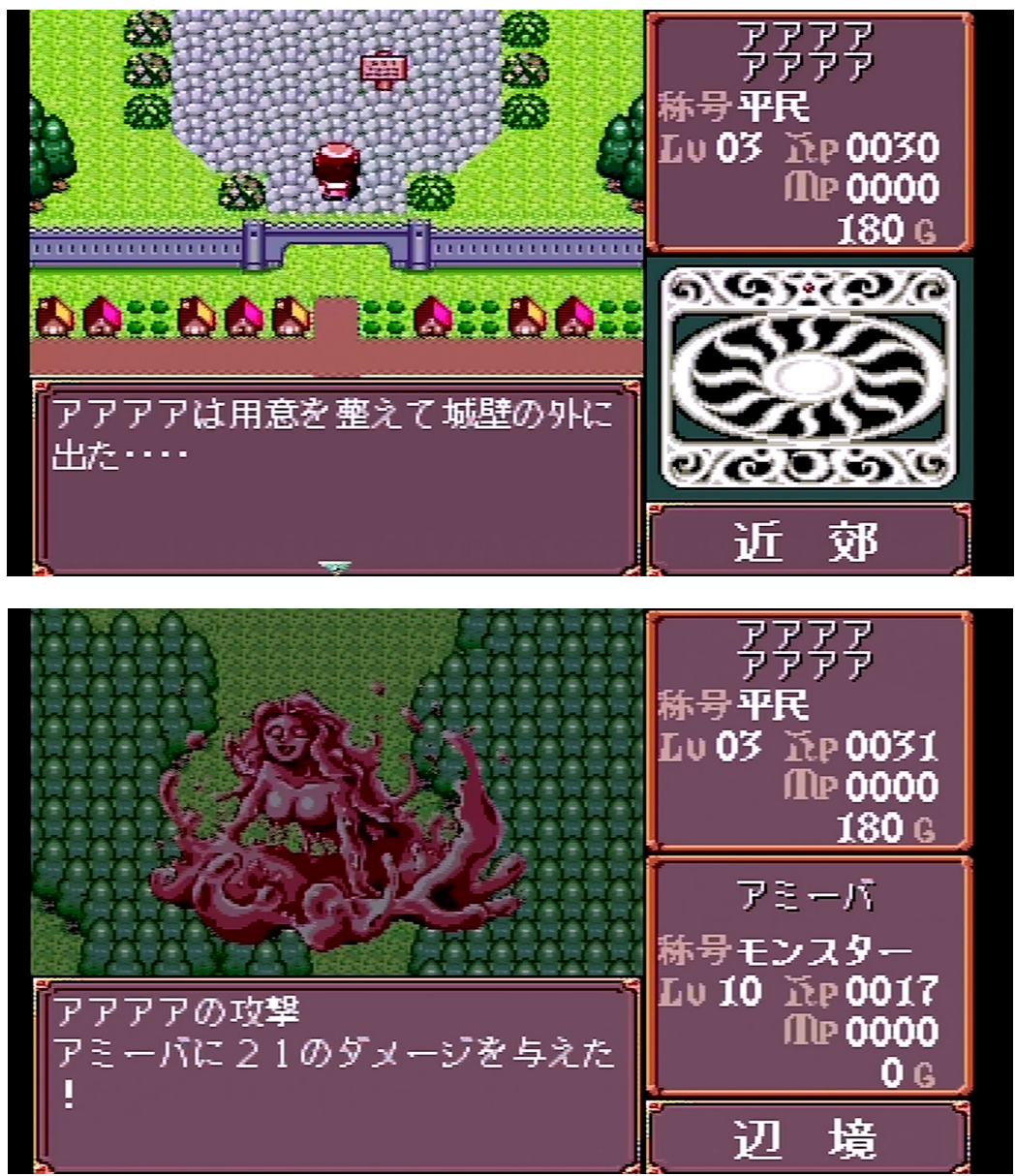

図 4-32：『プリンセスメーカー』における武者修行モード

『ダービースタリオン』では、（1）競走馬を育成するパートと、（2）育成した競走馬を実際にレースに出馬させるというパートが存在している。このゲームでは、育成パートキャラクター（競走馬）を育て、レースで良い成績を得ることが、ゲーム体験のより中心的な要素を占め、また、ゲーム内の中心的な目標となっており、この仕様を欠いてはこの作品自体の根幹

が成立しないと言ってもよい仕様となっている。（ただし、『ダービースタリオン』では、直接的にプレイヤーが操作をして介入するわけではない）

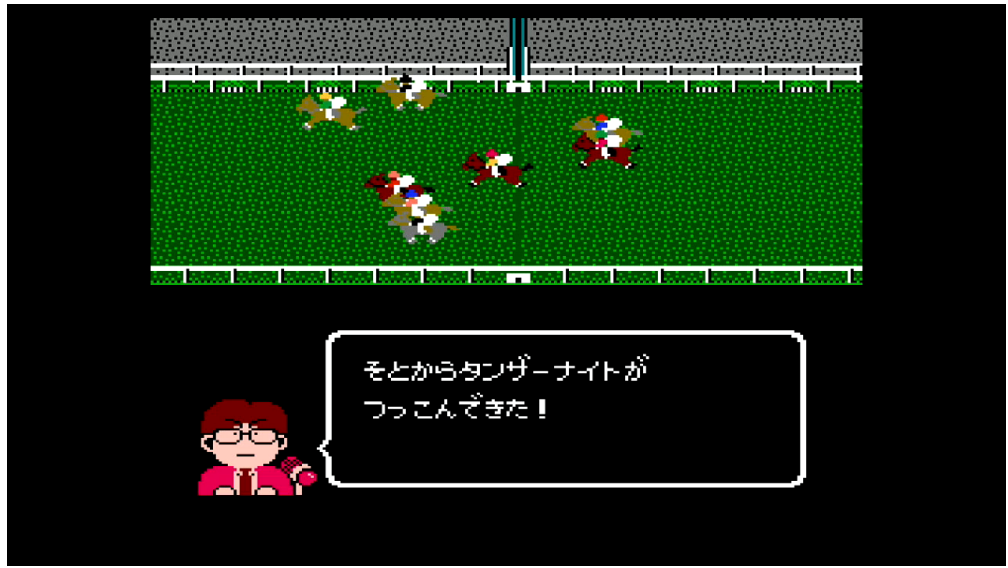

図 4-33：『ダビスタ』におけるレース出馬シーン

作品によって、何が「活用できるパート」に該当するのかには下記のようなバリエーションが見られるが、育成パートにおける各種パラメーターを成長させた成果が活用されるパートとしての機能をもっているという意味では、同様のゲーム仕様であると言える。

表 4-9：育成したキャラを活用できるパートの整理

| その作品が主に扱っているフィクションのテーマ | 活用モード | 作品例 |
| --- | --- | --- |
| スポーツチームの育成 | 試合 | 『サカつく』『パワプロ3』『パワプロアプリ』 |
| 中世ファンタジー風の世界 | ダンジョン探索／戦闘（RPGの仕様に準拠） | 『プリメ』『エタメロ』『悠久』 |
| 競走馬の育成 | レース | 『ダビスタ』『ウイポスワールド』 |
| ペットの育成 | 各種コンテスト | 『たまごっち』『nintendogs』 |
| モンスター等、戦闘能力を前提としたキャラクターの育成 | 対戦モード | 『モンファ』『ジョーカー2』『ポケモン ルビサファ』『艦これ』『とうらぶ』 |
| アイドルの育成 | ライブ | 『アイマス』 |

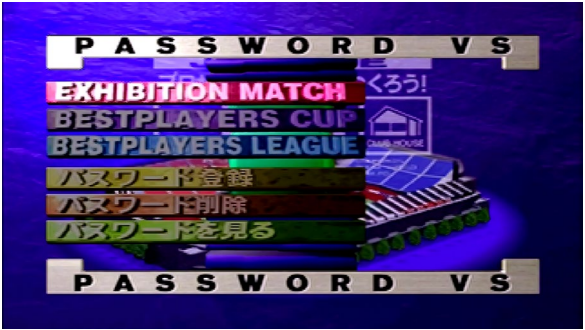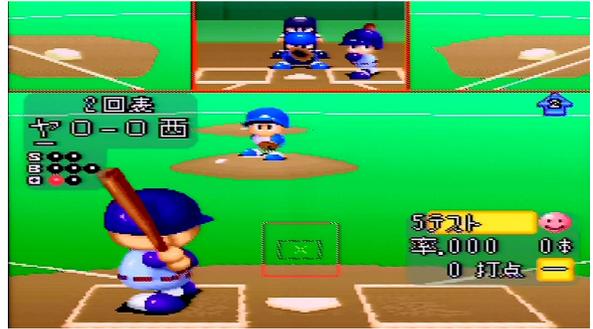

『サカつく』でも育成したサッカーチームで実際に試合をさせるモードが実装されている。

『パワプロ3』 サクセスモード

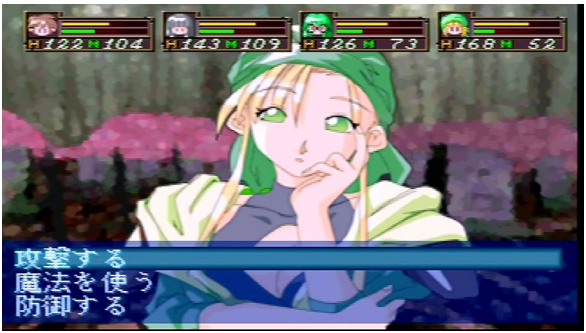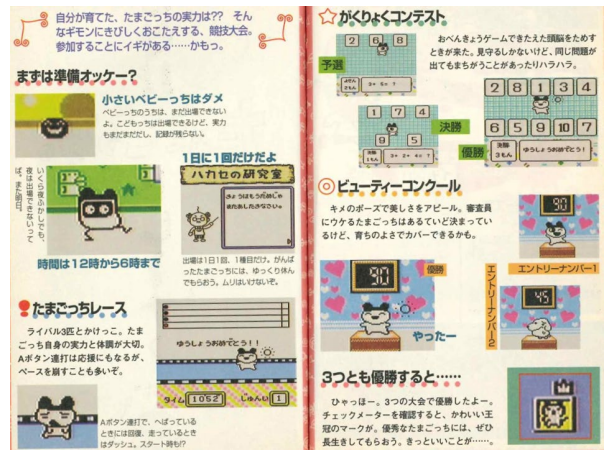

『エターナルメロディ』

『たまごっち』（アスペクト編集部 1997a, pp. 68-69）各種コンクール

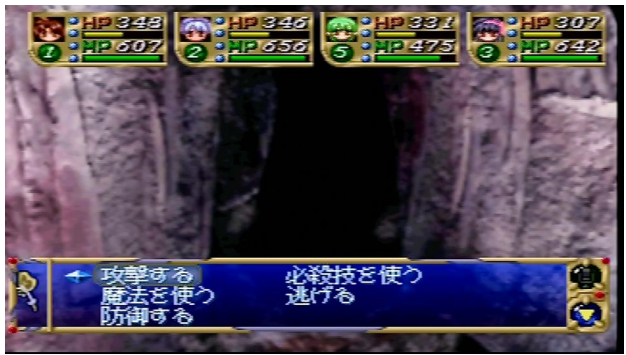

『悠久幻想曲』　ダンジョン

『モンスターファーム』、説明書 p.23

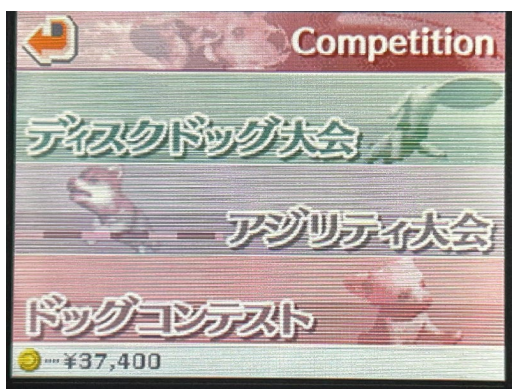

『nintendogs』

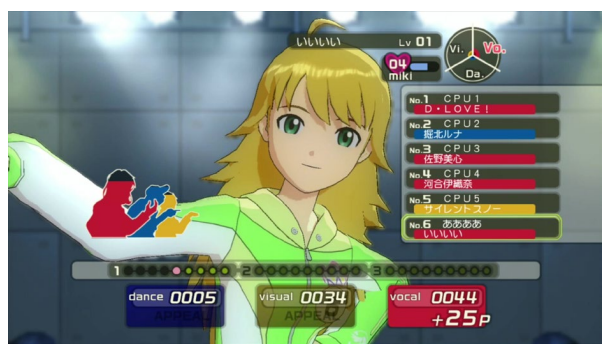

『アイドルマスター』

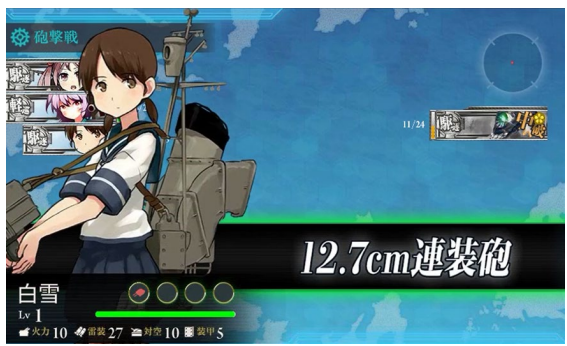

『艦これ』

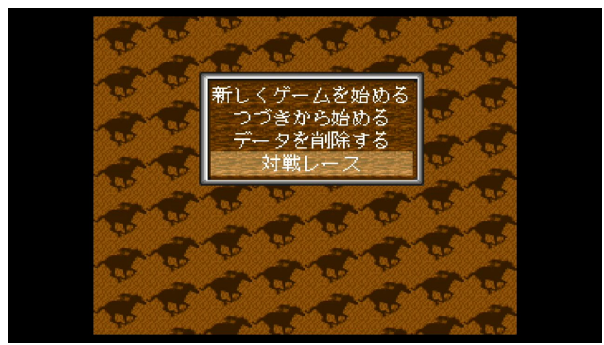

『ウイニングポスト 2』

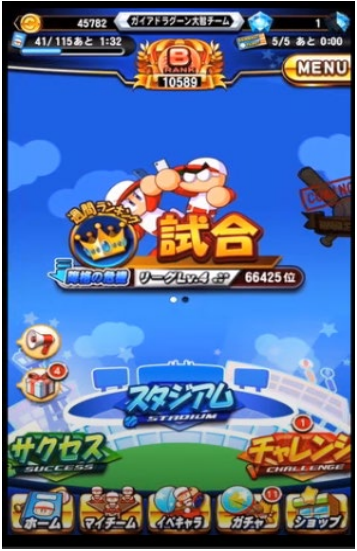

『パワプロアプリ』

図 4-34：その他のゲームにおける育成したキャラを活用できるパート

具体的な「活用パート」にあたるものの実装には作品ごとにバリエーションが見られるが、先述した通り「育成ゲーム」というジャンルが持つ性質上、この仕様はかなり初期から一般的に実装されているゲーム仕様の一つであると考えられる。

必須のゲーム仕様ではないが、典型的育成ゲーム群および「育成要素」を持つ作品の大半において実装されている仕様であると言えるだろう。

### 4.5.9 育成キャラと登場キャラの間の関係値に応じたイベント発生

育成ゲームは、ジャンルの発展当初から美少女キャラクターとの恋愛が、ゲーム内の要素に盛り込まれた作品が多く見られ、恋愛要素を持つ作品においては育成キャラと登場キャラの関係値に応じたイベント発生をゲーム仕様として盛り込むことが一般に確認できる。

たとえば、『卒業』は育成対象キャラクターらからの「人気度」というパラメーターが設定されており、この人気度のパラメーターがいくつであるかが、イベント発生の条件分岐のために用いられている。図 4-35 に、攻略本からの記述を示す。

**② ラブレターイベント発生条件**

①そのキャラクターのラブレターイベントがまだ発生していない
②日付が２月１日から２月19日の間
③「人気」パラメータが９０以上
④その日の課題が、生徒が自主的に選択した作文」で成功している

| イベント内容 | 発生条件 |
| --- | --- |
| アルバイト | 「魅力」が80以下の時発生しやすい。 |
| 夜の街 | 「人気」が50未満、かつ「品位」が50未満の時発生しやすい。 |
| 私服喫煙 | 「品位」が50未満の時発生しやすい。 |
| 制服喫煙 | 「品位」が30未満の時発生しやすい。 |
| 飲 酒 | 「異性」が80以上、かつ「品位」が45未満の時発生しやすい。 |
| 怪しいアルバイト | 「性欲」が90以上、かつ「品位」が45未満の時発生しやすい。 |
| 同 棲 | 「異性」が80以上、かつ「性欲」が80以上の時発生しやすい。 |
| ナンパ | 「異性」が90以上、かつ「品位」が80以上の時に発生しやすい。 |

図 4-35：『卒業』の関係値に応じたイベント発生条件 [27]

SegaSaturn magazine 編集部, アミューズメント書籍編集部編（1997, p.72）（上）

同前書（p.73）（下）

また、図 4-36 に示している通り、『ときめきメモリアル』は、男性主人公を魅力的な男性に育成することで、女性の好感度を上げることを主たる要素とした作品であるが、対象となる女

---

[27] ここでは『卒業 Graduation』ではなく『卒業 S』（イベント一部変更版）の攻略本を使用している。基本的な内容は PC エンジン版と同様とされている。

性（ここではメインヒロインの藤崎詩織[28]）からの主人公の評価が非常に低い状態（上）の場合、藤崎詩織からの反応が変わる（下）。

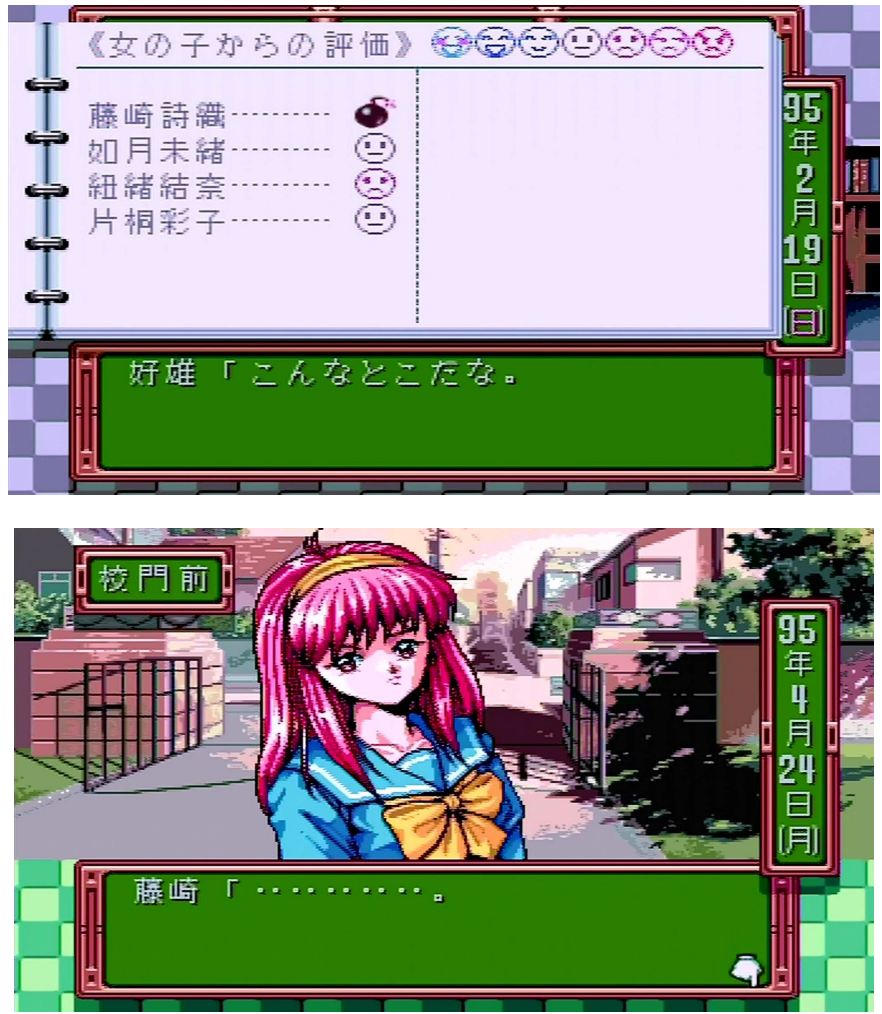

図 4-36：『ときメモ』におけるイベント発生

『エターナルメロディ』の攻略本（電撃 PlayStation＆電撃 SEGAEX 編集部, 1996, p.11）によれば、キャラクター同士の関係性を制御する値として「親近度」と「嫉妬度」が設定されてい

---

[28] 『ときめきメモリアル』における恋愛対象となる女性キャラクターである。藤崎詩織以外にも、複数の女性キャラクターが恋愛対象となっている。

る。嫉妬度について、「キャラクターたちはおたがいに対する嫉妬度を持っている。…（略）…嫉妬度が高いと喧嘩が発生する可能性がある。」と記述があり、キャラ同士の関係値に同じたイベント発生の仕組みが存在していることが示されている。

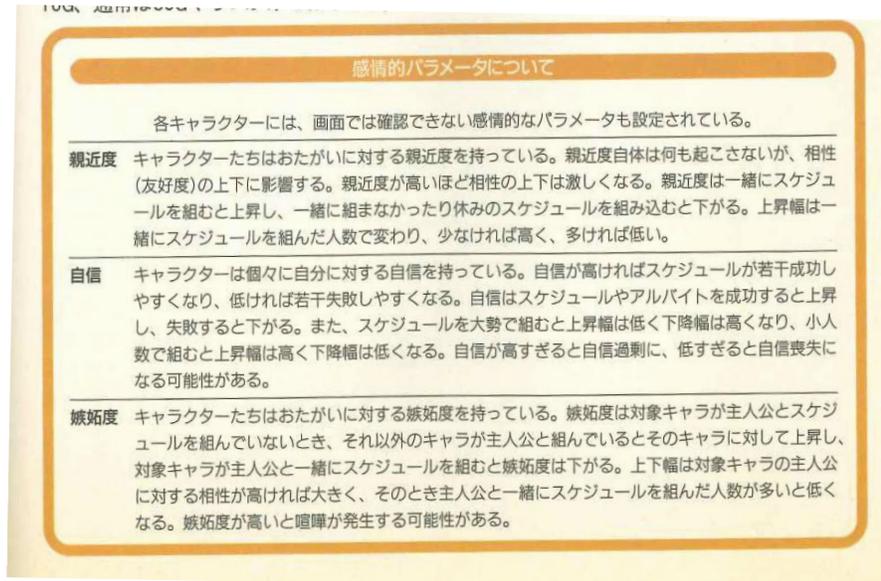

図 4-37：『エターナルメロディ』攻略本における各種関係値の説明

電撃 PlayStation＆電撃 SEGAEX 編集部(1996, p.11) より引用

　さらに、同書 p.75 によれば、主人公（育成キャラ）との相性が 55 以上か、75 以上かといった値によって、イベントの発生条件が分岐するように設定していることが示されている。

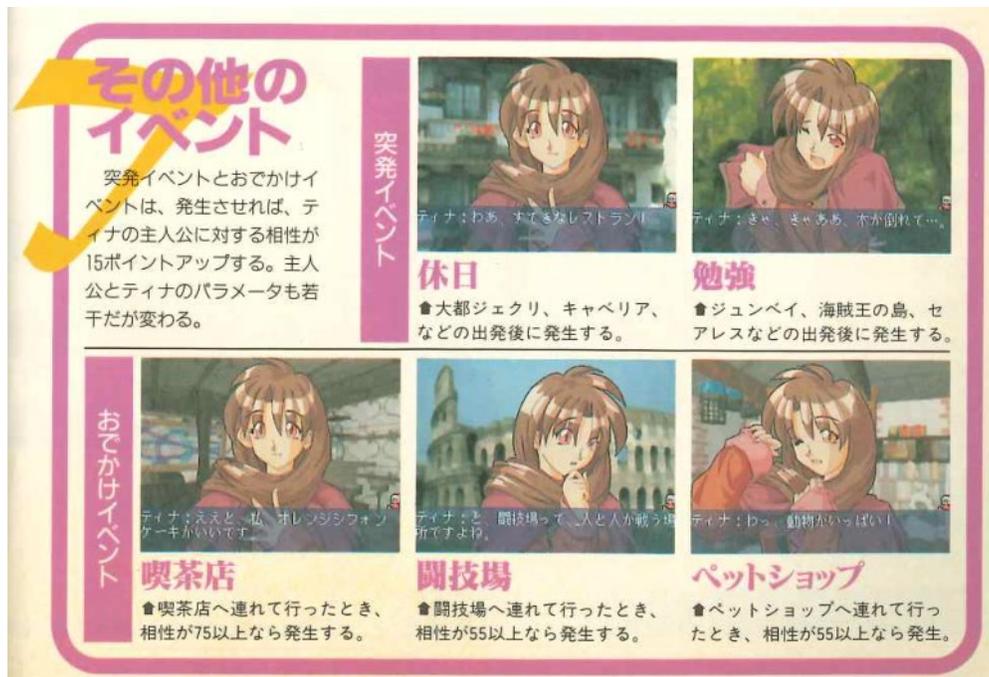

図 4-38：『エターナルメロディ』における関係値によるイベント分岐の説明

電撃 PlayStation＆電撃 SEGAEX 編集部(1996, p.75) より引用

　こういった育成キャラと登場キャラの間の関係値に応じたイベント発生の仕様は、育成要素およびキャラとの関係性といった仕様を有する作品であれば、育成ゲームとして典型的にみなされる作品でなくとも確認できる。

　例えば、育成ゲームではなく RPG である『ポケットモンスター ルビー・サファイア』には、育成しているモンスターの進化イベントの条件分岐において「なつき」度合いのパラメーターが参照されていることが示されている。

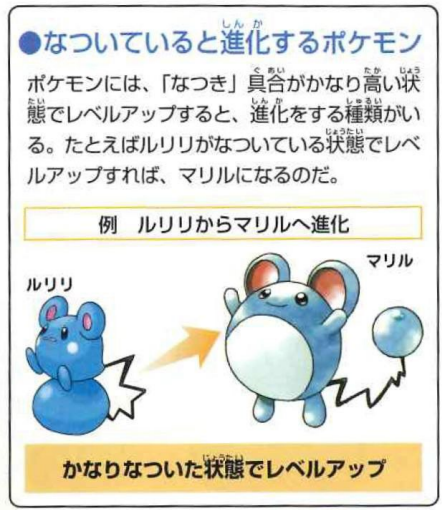

図 4-39:『ポケモン ルビー・サファイア』における関係値に応じたイベント

元宮秀介&ワンナップ編著（2002, p.19）

『nintendogs』においても同様に、「なつき」度でのイベント発生有無が制御されている。

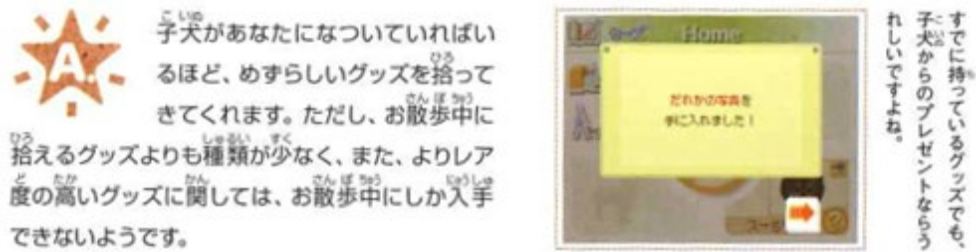

図 4-40:『nintendogs』における関係値に応じたイベント

ファミ通書籍編集部（2005, p.65）より引用

『パワプロアプリ』では、キャラとの関係値によってイベントの結果が影響を受ける。

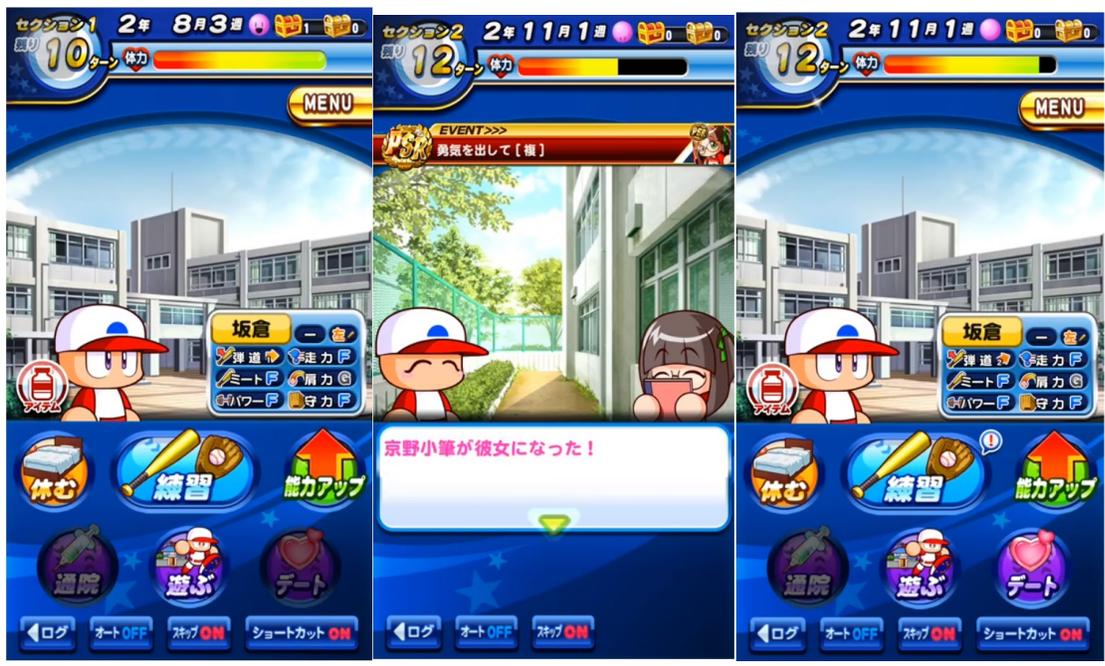

図 4-41：『パワプロアプリ』におけるキャラとの関係値に応じたイベント

関係値の上昇によって彼女ができ（真ん中）、デート項目が右下に追加された（右）

『FGO』では、育成キャラと登場キャラの間の「絆レベル」に応じて育成キャラに関するエピソードなどが読めるようになる。

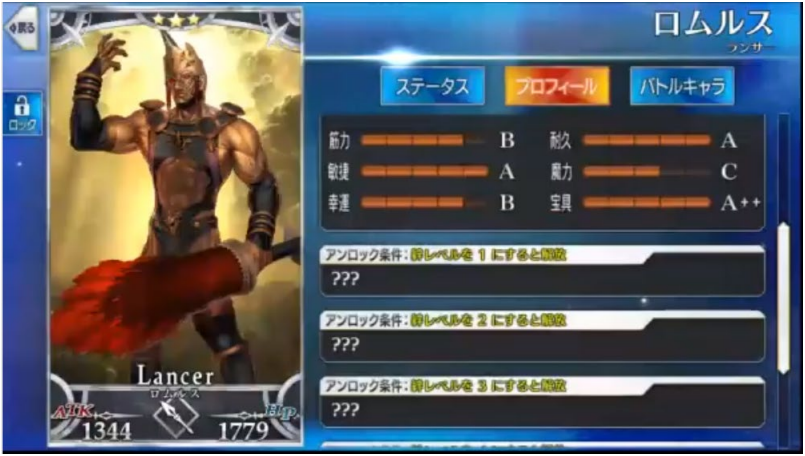

図 4-42：『FGO』におけるエピソード閲覧画面

以上のように、育成要素があるゲームにおいて、登場キャラと育成キャラとの間に何かしらの関係値（人気、なつき度、好意、友情、信頼度など）が設定され、当該関係値に応じたイベント発生をゲーム仕様として盛り込むことは、ごく一般的な仕様であると言うことができるだろう。

### 4.5.10 抽選で育成対象キャラクター等のゲーム内コンテンツが入手できる

ゲーム内における何らかの資源（お金、特殊な石など）を消費して抽選で育成対象キャラクター等のゲーム内コンテンツが入手できる仕様は、いわゆる「ガチャ」と呼ばれ、2010年代のスマートフォンやフィーチャーフォン等のゲームに広く見られるゲーム仕様である。海外のゲーム作品においてもルートボックス（Loot box）と呼ばれ、国内外のゲームに関わるニュースでもたびたび取り上げられる。

そのため、この時期のスマートフォン、フィーチャーフォンによるゲームを遊んだ多くのゲームプレイヤーや、SNS（Facebook, mixi, gree 等）を通したゲームを遊んだプレイヤーにとっては、ほとんどのゲームにおいて実装されていた仕様であるという感覚を持たれるものであろうと推測される。

実際、本調査において対象となった作品の中でも、ソーシャルゲームに類する全ての作品において、この特徴を確認することができた。

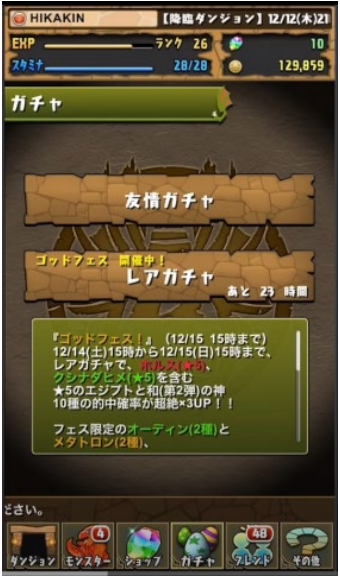

『パズドラ』

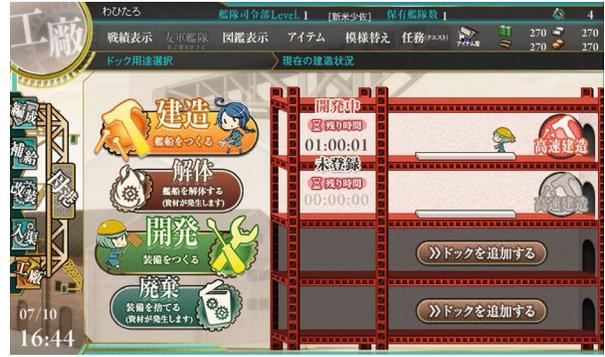

『艦これ』

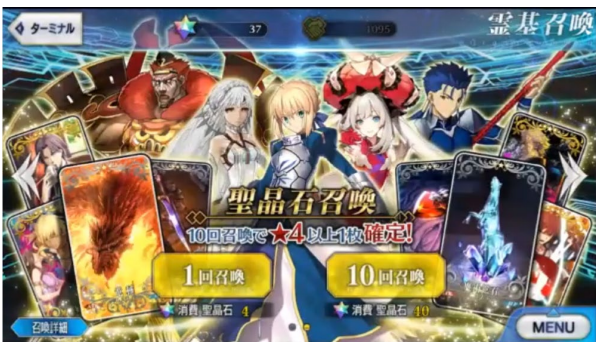

『FGO』

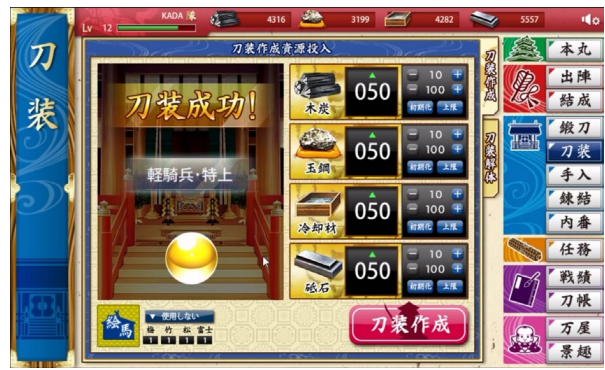

『刀剣乱舞』

図 4-43：各種ゲームにおける「ガチャ」

　この仕様は、育成ゲームであり、かつソーシャルゲームの仕様を持つ作品にも引き継がれており、『パワプロアプリ』においても、この仕様の存在を確認することができる。

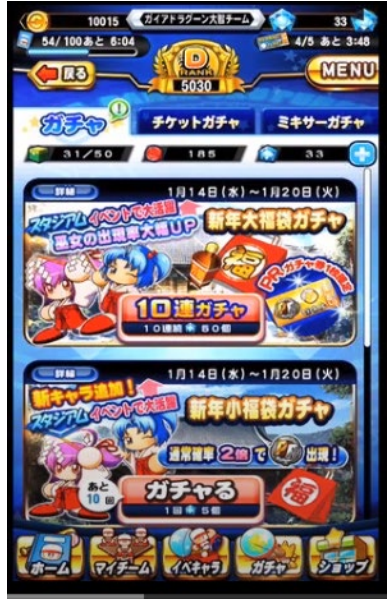

図 4-44：『パワプロアプリ』における「ガチャ」

なお、本調査の対象とした作品の中でもっとも初期にこの仕様が確認できたものとしては、2009 年の『ウイニングポストワールド』であった。

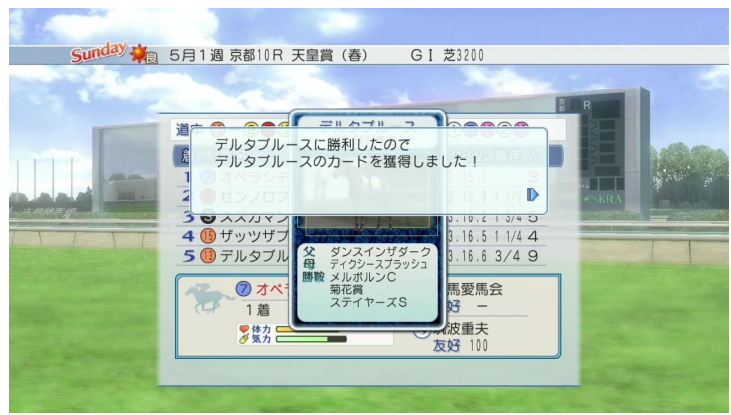

図 4-45：『ウイニングポストワールド』における「ガチャ」

育成ゲーム一般における仕様であるとは言えないが、2010年代以後のソーシャルゲームと呼ばれるほとんどのゲームにおいて確認することのできる仕様であり、育成ゲームジャンルにも流入してきた仕様であることが伺える。

### 4.5.11 育成対象キャラ等のゲーム内コンテンツにレアリティがある

4.5.10の仕様としばしばセットになっている仕様としてよく知られるものとして、育成対象キャラ等のゲーム内コンテンツに「ノーマル（N）」「レア（R）」「スーパーレア（SR）」といったレアリティが設定されている、ということが広く見られる。このレアリティには大きく分けて他の２つゲーム仕様と強く関わる形で実装されることが多い。

第一に、抽選の仕様(4.5.10)との関わっており、抽選における入手の希少性の差を示す指標となっている。入手困難なキャラであるほど「レアリティの高いキャラ」として抽選（ガチャ）において手に入りにくくなっている。

第二に、育成キャラを活用する仕様(4.5.8)と関わっており、レアリティの高いキャラほど育成対象キャラを活用する際に効果的なパラメーターの上限値が高く設定されている事が多い[29]。

レアリティの表記の仕方としてはいくつかのバリエーションがあるが、

（A）★の数によって希少性を示すもの

（B）ノーマル（N）、レア（R）、スーパーレア（SR）等の名称を記載するもの [30]

（C）アイコンによってレア度の違いが表現されているもの

---

[29] ただし、育成キャラを活用するモードがほとんど存在していないゲームにおいては、この第二の特徴は存在しないこともある。

[30] 他にコモン（C）、スーパーコモン（SC）など作品によってこの名称はやや異なっているが、多くが４～１０程度のレアリティの違いを設定している。

といった形の記載が多くものが見られる。特に、多く見られるのは（A）★による表記と（B）ノーマル、レア等による表記である。

（A）★によるレアリティの記載として、図 4-46 に『パズドラ』および『FGO』の例を示す。

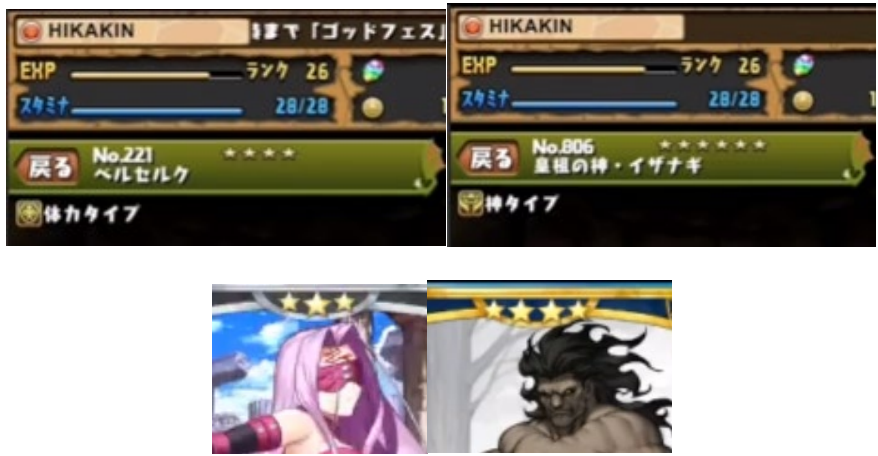

図 4-46：『パズドラ』（上）と『FGO』（下）における★によるレアリティ表記

（B）ノーマル、レア等の名称による表記として、図 4-47 に『パワプロアプリ』の例を示す。

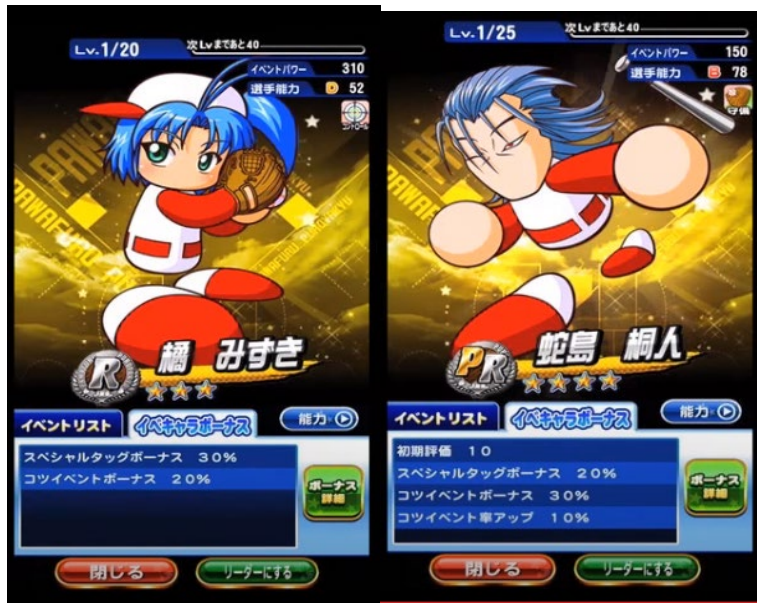

図 4-47：『パワプロアプリ』における名称によるレアリティ表記

R、PR の文字表記がレアリティを示す

（C）絵柄による表記として、図 4-48 に『艦これ』『刀剣乱舞』『なめこ』の例を示す

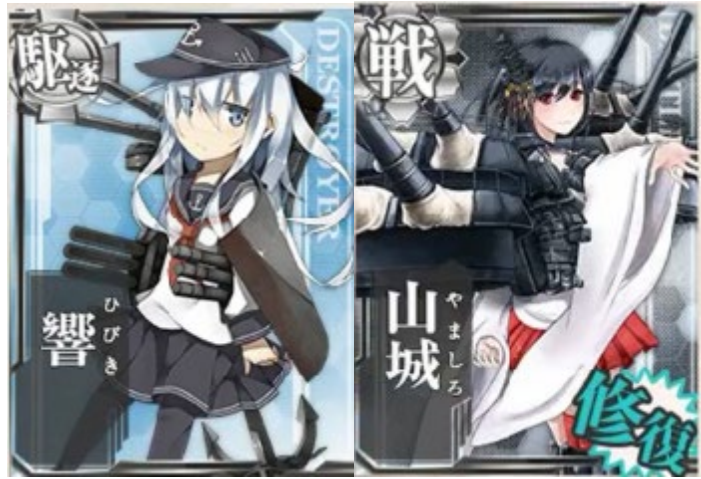

図 4-48：『艦これ』における絵柄による表記

『艦これ』では、背景色によってレアリティが異なっている

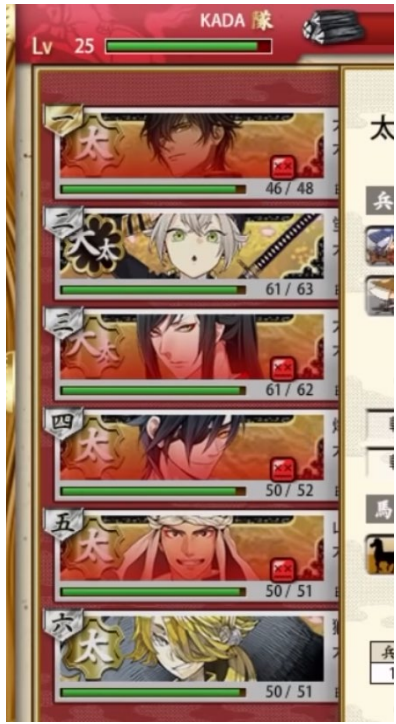

図 4-49：『刀剣乱舞』における絵柄による表記

キャラ左のアイコン（「太」「大太」の文字）の柄によってレアリティが区別されている。

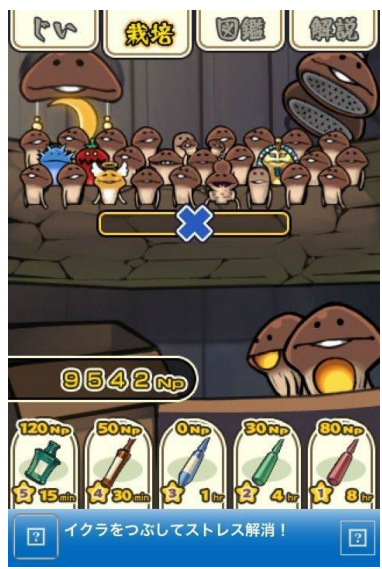

図 4-50：『なめこ』における絵柄による表記

羽の生えたなめこや、赤、青、黄色のなめこなど周囲と大きく異なる絵柄のもののレアリティが高い。

## 4.6 育成ゲームと他ジャンルとの合流について

すでに述べたように、育成ゲームは、育成ゲームジャンルの系譜内の作品のみではなく、他ジャンルとの間でゲーム仕様を継承・発展させながら発展してきている。その事例として、具体的に、育成ゲームの典型群とそれ以外の作品群のゲーム仕様が共通しているのかを示す。

まずロールプレイングゲーム群である『ポケットモンスター　ルビー・サファイア』および『ドラゴンクエスト IX』における仕様の共通性を示す。

図 4-51：『ポケットモンスター ルビー・サファイア』と他作品の仕様の共通度

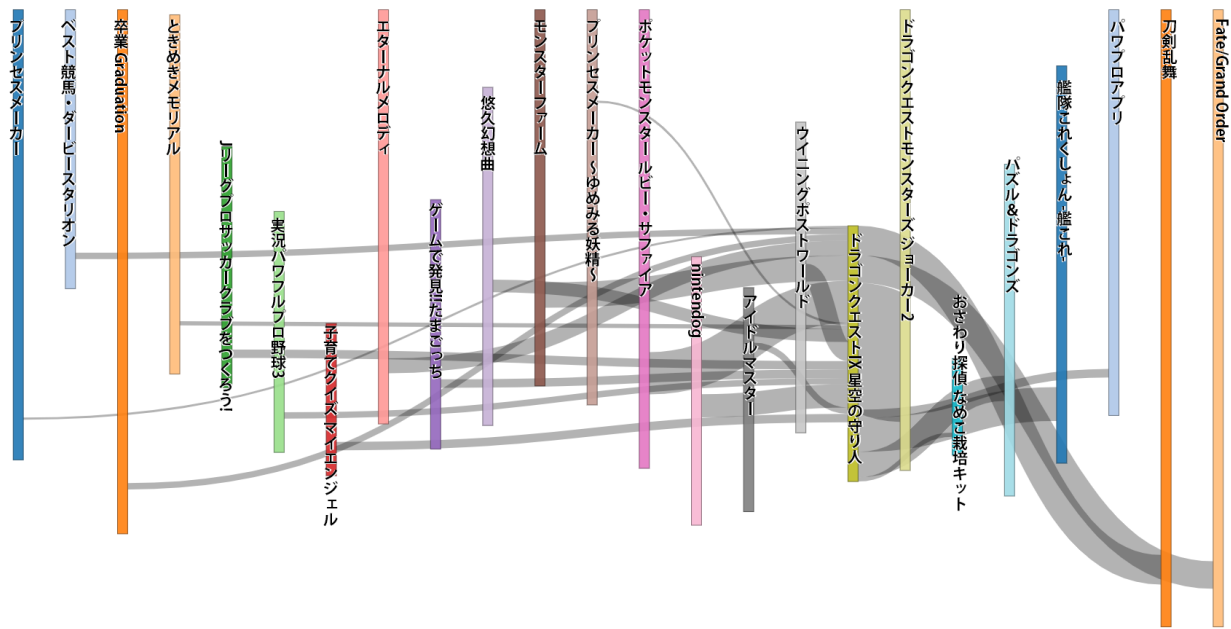

図 4-52 『ドラゴンクエスト IX』と他作品の仕様の共通度

図 4-51 と図 4-52 から明らかなとおり、先行する多くの育成ゲームと仕様が共通していること、また後発の多くの育成ゲームとも共通する仕様を有していることが伺える。

次にソーシャルゲームである『パズル＆ドラゴンズ』と他作品の仕様の共通度について示す。

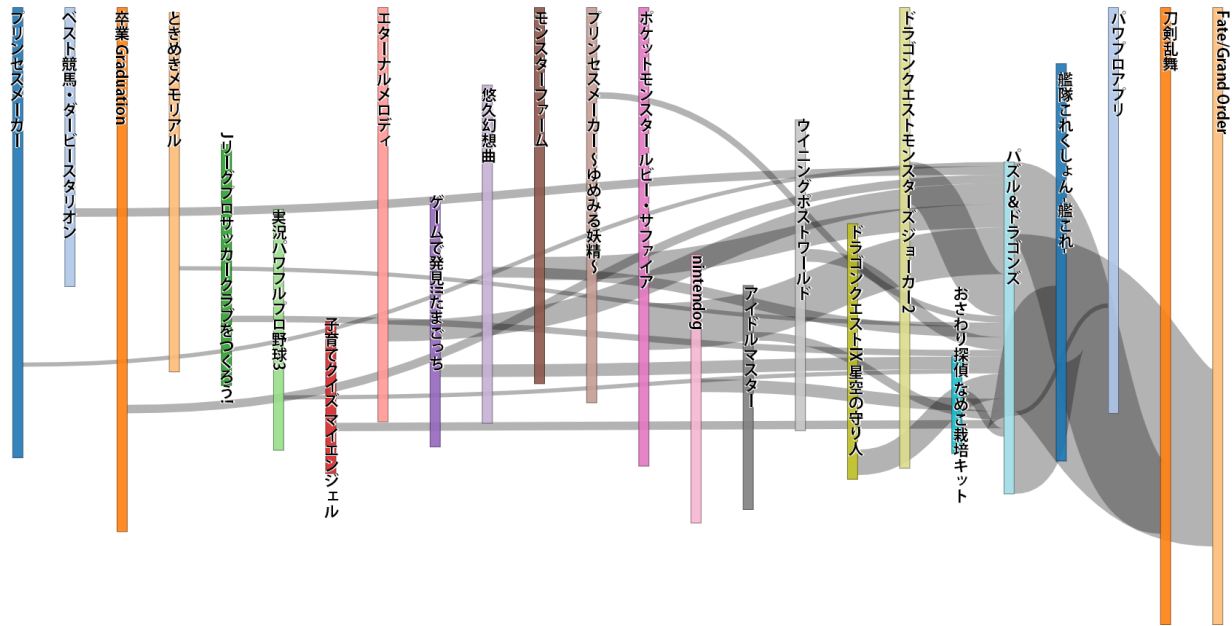

図 4-53『パズル＆ドラゴンズ』と他作品の仕様の共通度

図 4-53 が示すとおり、他の作品の仕様を継承していると同時に、他のソーシャルゲーム群である『艦これ』、『刀剣乱舞』、『FGO』へ影響すると同時に、育成ゲームでありかつソーシャルゲームである『パワプロアプリ』とも共通性を有していることがわかる。

## 4.7 小括：育成ゲームジャンルの概要と系譜について

以上、本手法を通じて育成ゲームジャンルについて次のような点を示した。

- 育成ゲームのジャンル概念が 1990 年代から発展したものであること（4.1）
- 育成ゲームの作品群が相互に多くの要素を共通した要素として持っていること(4.2)
- 育成ゲームの作品群の中で特に典型的な作品とその準典型作品群、隣接領域の作品群などのクラスターの分割が説得的に示すことができること(4.3)
- 育成ゲームにとりわけ特徴的なゲーム仕様として複数の育成コマンドがあり、決められた範囲の育成期間が設定されていることなどが挙げられること（4.4）
- 具体的なゲーム仕様の実現方法にはさらに様々な形式が見られること(4.5)

- 育成ゲームの作品群が、育成ゲーム以外のジャンルとも相互にゲーム仕様を影響させつつ継承・発展してきたこと(4.6)

## 5.結論：本手法の評価と検討

### 5.1 本手法の新規性：標準化、再現性、効率性

　本手法の最大のメリットは、属人的な知識に可能な限り頼らない形での、ジャンル内作品の内容の近さを示すとともに、その内容の系譜記述が可能な方法を示したことである。

　可能な限り、データを定量的に取り扱える形にしたことで、作品間のゲーム仕様に基づいた内容の一致率を示すための手法を提案できたものと考える。これにより、第三者により類似のデータセットを作成し、本研究で示された内容について再現性を検討できるような形式を提示できたものと考える。

　また、具体的なゲーム仕様を列挙した上で、そのゲーム仕様間の類似性を判定する手法は、文の文言上の一致率を確認するのみであれば、機械学習以前から存在するレーベンシュタイン距離(Levenshtein 1966)を用いた方法などもあったが、BERT のような高度な自然言語処理を本調査研究において用いたことで、効率的に類似性を示すことができるようになっている [31]。

### 5.2 本手法の限界：起源の同定(影響の明確な因果関係)、体験の全体像の素描、

　本手法において達成できていない点は大きく分けて 2 点ある。

　第一に、 本手法では特定のゲーム仕様の最初の起源が、明確にいつであったかの同定を行うことはできない。これは、本手法の限界でもあるが、標準化された手法によって「起源」を明らかにすることは、そもそも容易なことではない。考えうる手法の一つとしては、初期の作品

---

[31] 機械学習を用いたゲームのジャンル分類への応用については、大山ら(2018)などがあるが、ゲームの言説の検討および、人間による直接のゲームの内容分析と組み合わせて統合的に分析した研究としては本研究に新規性はあるものと考える。

とみなされている作者の開発者にインタビュー等を実施し、作品製作にあたって、影響を受けた作品があるかといったことを芋づる式に聞いていくという手法が想定されるが、決して効率的な手法ではない。

ただし、特定のゲーム仕様の起源や、作品間の直接の因果的な影響関係についてまで明らかにすることはできなくとも、どの時点においてすでにゲーム仕様がどの程度広く存在しているかといった関係を確認するためのデータについては本手法を通じて示すことができる。本手法は起源や影響関係の最終的な同定はできないものの、その同定プロセスに貢献することはできる。

第二に、個別のゲーム仕様の要素に還元しえないゲームの体験の全体像の分析は、本手法では明らかにならない。Arsenault (2009, p.171) は、「ゲームのジャンルは、抽象化されたチェックリストのような機能ものはなく、ゲームプレイ体験を通じての行動の現象学的、実践的なありようを形成しているものである。」し、チェックリスト方式の分析を批判している。ゲームの体験には、個別の要素に還元不可能な要素が存在するという批判自体は妥当な批判であると言える。この「総体的な体験」の分析を目指すのであれば、本手法ではなく GTA などを用いた質的な体験の分析手法や、より抽象度の高い次元の分析（Yee 2006 など）が考えられる。

## 6.手法の標準化のための改善について

今回用いた BERT は育成ゲームに関連する概念を学習させる等のチューニングを行っていない。該当する文章を予め学習させることで、自然言語処理による類似度の判定精度はより高くなることが期待される。

## 7.謝辞

# 文献リスト